\def\version{30 July 1999}%
\def\preprint{DAMTP-1999-93}%

\def\tr{\mathop{\rm tr}\nolimits}
\def\ad{\mathop{\rm ad}\nolimits}
\def\rk{\mathop{\rm rank}\nolimits}
\def\diag{\mathop{\rm diag}\nolimits}

\def\nn{\nonumber}%
\def\g{{\frak g}}
\def\N{{\Bbb N}}
\def\C{{\Bbb C}}
\def\su{{su}}
\def\so{{so}}
\def\sp{{sp}}
\def\ie{{\sl i.e.\/}}
\def\eg{{\sl e.g.\/}}

\def\etc{{\sl etc.\/}}
\def\cf{{\sl cf.\/}}

\def\pacs#1{}%
\def\keywords#1{}%

\def\slantfrac#1#2{{#1}/{#2}}%
\def\openone{\leavevmode\hbox{\small1\kern-3.8pt\normalsize1}}%

\def\acknowledgements{\section*{Acknowledgements}}%
\def\address#1{\date{{\sl#1}\\\ \\\version}\gdef\date##1{}}%
\def\hhline{\hline\hline}%
\def\dopreprint{\hfill{\small\preprint}\\}%

\def\mycite#1{\cite{#1}}%
\def\Mycite#1{Ref.~\cite{#1}}%
\def\mycites#1{\cite{#1}}%

\documentstyle[cite,11pt,twoside,a4wide,amsfonts]{article} 

\renewcommand{\theequation}{\thesection.\arabic{equation}}%
\makeatletter
\@addtoreset{equation}{section}
\makeatother

\newcounter{mathletter}%
\newcommand{\bmathletter}{%
  \refstepcounter{equation}%
  \setcounter{mathletter}{\value{equation}}%
  \setcounter{equation}{0}%
  \renewcommand{\theequation}{%
    \mbox{\thesection.\arabic{mathletter}\alph{equation}}}}
\newcommand{\emathletter}{\setcounter{equation}{\value{mathletter}}}%
\newenvironment{mathletters}{\bmathletter}{\emathletter}%

\def\begintab#1{\begin{tabular*}{\textwidth}{@{\extracolsep\fill}#1}}%
\def\endtab{\end{tabular*}}%
\def\beforetab{\renewcommand{\arraystretch}{1.25}}%
\def\aftertab{\renewcommand{\arraystretch}{1}}%
\begin{document}
%

\title{\dopreprint On characteristic equations, trace identities and
  Casimir operators of simple Lie algebras} 
\author{A.~J.~Macfarlane\thanks{e-mail: A.J.Macfarlane@damtp.cam.ac.uk}\ \ 
  and Hendryk Pfeiffer\thanks{e-mail: H.Pfeiffer@damtp.cam.ac.uk}}
\address{Department of Applied Mathematics and Theoretical Physics,\\
  Silver Street,\\ Cambridge CB3 9EW} 
\date{\version}
\maketitle

%
\begin{abstract}
%
  
  Two approaches are developed to exploit, for simple complex or compact real
  Lie algebras $\g$, the information that stems from the characteristic
  equations of representation matrices and Casimir operators. These approaches
  are selected so as to be viable not only for `small' Lie algebras and
  suitable for treatment by computer algebra. A very large body of new results
  emerges in the forms, a) of identities of a tensorial nature, involving
  structure constants \etc\ of $\g$, b) of trace identities for powers of
  matrices of the adjoint and defining representations of $\g$, c) of
  expressions of non-primitive Casimir operators of $\g$ in terms of primitive
  ones. The methods are sufficiently tractable to allow not only explicit
  proof by hand of the non-primitive nature of the quartic Casimir of $g_2$,
  $f_4$, $e_6$, but also \eg\ of that of the tenth order Casimir of $f_4$.

\end{abstract}


%
\section{Introduction}
%

This paper is concerned with two related matters. One is the analysis of
higher order Casimir operators of a simple complex or compact real Lie algebra
$\g$ in terms of primitive Casimirs. The other is the provision of identities
involving the structure constants of $\g$ and related invariant tensors, such
identities being often presented as expressions for the trace of the product
of matrices from either the defining or the adjoint representation of $\g$.

Much of course is known about these matters, and we attend with care below in
the last paragraphs of this introduction to the relationship of our work to
that of previous authors. Our purpose however has been to develop and apply
methods that remain viable for $\g$ of large rank or dimension, and that are
amenable to treatment by computer algebra. In fact, we claim to have a large
body of new results, for many Lie algebras, not only for $g_2$, $f_4$ and
$e_6$, but also for the classical families, with even quite a few for
$a_2=\su(3)$.

It is well known (see~\mycites{MaSu68,AzMa98}, for example, for
$a_\ell=\su(\ell+1)$) that identities for ad-invariant tensors for $\g$ can be
separated into two classes. The first class contains results readily available
for all $\g$, and having a common appearance for all $\g$ in any of the four
classical families, which arise (as section~\ref{sect_sun} below explains in
detail) from use of Jacobi type identities. The second class of identities,
which take on a form special for each Lie algebra $\g$, arise from the use in
some way of the characteristic equations of representation matrices and
Casimir operators. The paper aims to justify its existence by its approaches
to identities in the second class, to the results that flow from them, and
from their application to the treatment of non-primitive Casimir operators.

We follow here two approaches to class~2 identities and characteristic
equations. One is based on the characteristic equation of the second order
Casimir operator on $V_{\rm adj}\otimes V_{\rm adj}$ where $V_{\rm adj}$ is
the adjoint representation of $\g$. The other proceeds by diagonalization of
representation matrices of the defining and adjoint representations of $\g$.

It is convenient to explain these methods first for $a_2=\su(3)$ or, at some
points, $\su(n)$. Thus our paper is organized as follows. After a few brief
remarks in section~\ref{sect_sun} about notation, we take up our two methods
for $\su(3)$ in sections~\ref{subsect_methodproj} and~\ref{subsect_traces}.
While often the results here are well-known, their derivations may well be
simpler, and, as subsequent work indicates, more amenable to generalization
than previous ones. Furthermore, one finds, even in this well-studied context,
new results of clear importance (\cf~(\ref{resf6_a2}) and~(\ref{resf8_a2i})).

Section~\ref{sect_appl} then proceeds briefly through application of the first
method to the four classical families of Lie algebras and to $g_2$, $f_4$ and
$e_6$, displaying a wide variety of new results, especially for the three
exceptional algebras.  Our strength ran out, as did the calls made on
information storage by our computer programs, during the study of $e_6$, so
that $e_7$ and $e_8$ have not been treated. Since our second method requires
little explanation beyond that given in section~\ref{subsect_traces} for
$\su(3)$, and yields useful output easily by hand for simple cases, we simply
present the results in an appendix. Much here is new and we think valuable.

For what is merely a selection of interesting items, we refer to
eqs.~(\ref{tracea2_fa}) and~(\ref{tracea2_ff}) for $a_2=\su(3)$, to the
situation surrounding the primitivity for $a_4=\su(5)$ of the traces $\tr
F^{2k}$, to~(\ref{tracee6_fa}) for $e_6$, (\ref{tracef4_ff}) for $f_4$ and
corresponding results for $g_2$. As an indication of results for which pen and
paper derivation were viable ahead of their confirmation by algebraic
computation, we cite, for example, (\ref{traceg2_faa})--(\ref{traceg2_fad})
and~(\ref{traceg2_ffa}) and~(\ref{traceg2_ffb}), as well
as~(\ref{tracef4_aab}), yes, for $f_4$.

A C-program was used to produce data regarding representations of $\g$ and
the reduction of tensor products with only the Cartan matrix as input. Maple
was used to prepare the appendix.

Finally, attention must be given to placing our paper in the context of
previous work. To begin this, we recall that work on $\su(3)$ tensorial
quantities, the two classes of identities, and use of characteristic
equations, began many years ago, see
\eg~\mycites{MaSu68,AzMa98,Su69,Su90,RaSa73} for $\su(n)$. See
also~\mycite{Su69,Su90} for class~1 identities. The study of Casimir
operators~\cite{Ra50,Ra51,Ra65} likewise has a long history, which may
be further traced from~\mycite{AzMa98}: see references 2 to 8
in~\mycite{AzMa98}. That paper, \mycite{AzMa98}, contains a canonical
definition via cocycles \cite{AzPe97,AzIz95} of a set of primitive
Casimir operators for $\g$. It also relates these to a set of totally
symmetric isotropic tensors for $\g$. In the process, various
identities for such tensors are formed, especially for $\su(n)$. The
present paper reproduces many of these, by methods that are in general
easier and more readily extended. \Mycite{AzMa98} employed the
identities featured in it in the reduction of non-primitive Casimir
operators in terms of primitive ones, a topic also carried much
further here.

Next, we comment on the non-primitive nature of the quartic Casimir of $g_2$,
proved by Okubo~\cite{Ok77} on the basis of identities whose proof was not
displayed by him, although it needs only an easy calculation especially using
the method of section~\ref{subsect_traces}. Meyberg~\cite{Me83} gave a proof
of a similar result for all exceptional Lie algebras, and Cvitanovi{\v c}
indicated (only, as far as we know, via private communication to Okubo) that
proof of the same result was available via `bird-track' methods~\cite{Cv84}.
Our method of proof was easy enough to enable us to extend it by hand to show
the non-primitive nature of the tenth order Casimir of $f_4$.

It should be remarked also there is much common ground between the outlook of
this paper and the work of Cvitanovi{\v c}~\cite{Cv84,Cv77} even though the
latter is devoted to the development of diagrammatic methods.

Next, we note that Meyberg~\cite{Me93} used the decomposition of $V_{\rm
  adj}\otimes V_{\rm adj}$ in conjunction with trace identities, but only as
far as fourth powers, and not in conjunction with tensorial methods. Another
significant work on trace calculations is that of Mountain~\cite{Mo98}. His
method naturally covers very many cases where no special identities, like
those which reflect the absence of the quartic Casimir of $g_2$, are present,
and is effective also when supplemented by such identities where necessary.
The work of Gould~\cite{Go85} discusses invariant polynomials and
characteristic equations in a formulation using the universal enveloping
algebra. There is further information also in the earlier
works~\mycite{BrGr71,Gr71}.

We also cite the Physics Report of Slansky~\cite{Sl81}, which contains many
tables of Lie algebra data useful in the present work. Likewise useful for
background information are the books~\cite{AzPe97,FuSch97} and the
paper~\cite{Ok79}.

%
\section{Notation and conventions}
%
\label{sect_sun}

Let $\g$ be a simple complex or compact real Lie algebra. For the general
discussions we choose a basis $(X_j)$ of $\g$ such that the Cartan-Killing
form is $\kappa_{jk}=\tr(\ad X_j\circ\ad X_k)=-\delta_{jk}$, and write the Lie
product, with totally antisymmetric structure constants $C_{jk\ell}$, in the
form
\begin{equation}
\label{lie_bracket}
  [X_j,X_k]=C_{jk\ell} X_\ell\,.
\end{equation}

In discussing specific examples we adhere to conventions commonly used in the
physics literature. Thus, for $\su(n)$, we use the Gell-Mann matrices
$\lambda_j$. These are a set of $n\times n$ traceless hermitean matrices,
normalized according to $\tr(\lambda_j\lambda_k)=2\delta_{jk}$. They have the
multiplication law
\begin{equation}
\label{gell-mann}
  \lambda_j\lambda_k = \frac{2}{n}\delta_{jk}\openone+(d_{jk\ell}+if_{jk\ell})\lambda_\ell,
\end{equation}
where the completely symmetric $d$-tensor satisfies \cite{MaSu68,Su90}
$d_{jj\ell}=0$. Since (\ref{gell-mann}) implies
$[\lambda_j,\lambda_k]=2i\,f_{jk\ell}\lambda_\ell$, and the basis vectors of
$\su(n)$ are defined by $x_j=\slantfrac{\lambda_j}{2}$, the $f_{jk\ell}$
serve as structure constants for $\su(n)$. The relationship of this basis to
that of the general discussion is seen via $\tr(\ad x_j\circ\ad
x_k)=n\,\delta_{jk}$, and is given by $X_j=\slantfrac{i\,x_j}{\sqrt{n}}$ and
therefore $C_{jk\ell}=-\slantfrac{f_{jk\ell}}{\sqrt{n}}$.

The well known class~1 identities which are valid in general for all $n$
follow from Jacobi type identities
\begin{mathletters}
\begin{eqnarray}
0 &=& [[\lambda_j,\lambda_k],\lambda_\ell]+[[\lambda_k,\lambda_\ell],\lambda_j]
      + [[\lambda_\ell,\lambda_j],\lambda_k],\\
\label{jacobi_mixed}
0 &=& [[\lambda_j,\lambda_k],\lambda_\ell]+\{\{\lambda_j,\lambda_\ell\},\lambda_k\}
      - \{\{\lambda_k,\lambda_\ell\},\lambda_j\},\\
0 &=& [\{\lambda_j,\lambda_k\},\lambda_\ell]+[\{\lambda_k,\lambda_\ell\},\lambda_j] 
      + [\{\lambda_\ell,\lambda_j\},\lambda_k],
\end{eqnarray}%
\end{mathletters}%
together with consequences based on trace properties and completeness
relations for the $\lambda_j$. Class~2 identities emerge in forms different
for each $n$. Many of them are absent or degenerate for $n=2$, and, in
general, exhibit a complexity that increases with $n$. Such class~2 identities
stem from the use of the characteristic equation~\cite{MaSu68,Su90}, \eg\ of
$A=a_j\lambda_j$, $a_j\in\C$. The procedure is already
cumbersome~\cite{AzMa98,Su69,Su90,RaSa73} for modestly large $n$.

%
\section{Description of the methods}
%
\label{sect_method}

In this section we describe our methods and apply them to $\su(n)$, $n=3$ or
$4$ for illustrative purpose.

\subsection{Tensor products of the adjoint representation}
\label{subsect_methodproj}

Our first method uses the tensor product $V_{\rm adj}\otimes V_{\rm adj}$,
where $V_{\rm adj}$ denotes the adjoint representation of $\g$, and is based
on the characteristic equation of the second order Casimir operator. It offers
a convenient way of deriving certain class~2 identities, and as a by-product
yields explicit expressions (in tensorial form) for Clebsch-Gordan
coefficients occurring in the reduction.

Consider the tensor product of a finite-dimensional irreducible representation
$V$ of $\g$ and its decomposition into irreducible components $W_i$
\begin{equation}
\label{decomposition}
  V\otimes V\simeq W_1\oplus W_2\oplus\cdots\oplus W_k\,.
\end{equation}
For example let $V$ be the adjoint representation $[8]$ of $\su(3)$, then
\begin{equation}
\label{decomposition_a2}
  [8]\otimes[8] = \underbrace{[1]\oplus[8]\oplus[27]}_{\rm symmetric}
    \oplus \underbrace{[8]\oplus[10]\oplus[\overline{10}]}_{\rm antisymmetric},
\end{equation}
or in $\su(4)$
\begin{equation}
\label{decomposition_a3}
  [15]\otimes[15] = \underbrace{[1]\oplus[15]\oplus[20]\oplus[84]}_{\rm symmetric}
    \oplus \underbrace{[15]\oplus[45]\oplus[\overline{45}]}_{\rm antisymmetric}.
\end{equation}
These decompositions are always understood over the field of complex numbers.
In the basis $(X_i)$, the second order Casimir operator on $V$ is given by
\begin{mathletters}
\label{casimir_def}
\begin{equation}
  C_V = -\sum_{r=1}^n X_r^2\,.
\end{equation}
It acts on $V\otimes V$ as
\begin{equation}
  C_{V\otimes V} = -\sum_{r=1}^n{(X_r\otimes\openone+\openone\otimes
  X_r)}^2 = C_V\otimes\openone + \openone\otimes C_V+2\,L,
\end{equation}
where we have defined
\begin{equation}
  L=\frac{1}{2}(C_{V\otimes V}-C_V\otimes\openone-\openone\otimes C_V)
   =-\sum_{r=1}^n X_r\otimes X_r\,.
\end{equation}%
\end{mathletters}%
$L$ has the same $\g$-invariant subspaces as $C_{V\otimes V}$. Written with
indices, it has the form
\begin{equation}
\label{def_l}
  L_{jk,pq} = -C_{jpr}C_{kqr}\,.
\end{equation}

The normalization of the Casimir operators obviously is important when their
eigenvalues are considered. Our normalization of $C_V$ is such that the
adjoint representation has eigenvalue $1$ regardless of the algebra. The
eigenvalue is furthermore equal to $\left<\Lambda,\Lambda+2\,\delta\right>$
where $\Lambda$ is the highest weight of a finite-dimensional irreducible
representation, $\delta$ denotes the half-sum of positive roots of $\g$ and
$\left<\cdot,\cdot\right>$ is induced from the Cartan-Killing form on the
space of weights.

When $V$ denotes the adjoint representation of $\su(n)$, we have, for example,
$L_{jk,pq}=-\frac{1}{n}f_{rpj}f_{rqk}$. If we were working only with
$\su(n)$, we would absorb the factor $\slantfrac{1}{n}$ into the definition of
$L$ to make the eigenvalues integral, but this does not allow uniform
treatment of algebras of different series.

Assume that in the decomposition (\ref{decomposition}) $m$ projectors
$P^{(1)},\ldots,P^{(m)}$ onto the $W_1,\ldots,W_m$ are explicitly known
($m<k$). Then $P^{({\rm others})}=\openone-P^{(1)}-\cdots-P^{(m)}$ projects
onto the sum $W_{m+1}\oplus\cdots\oplus W_k$. The characteristic equation of
$L$ (which has the same structure as that of $C_{V\otimes V}$) implies
\begin{equation}
\label{characteristic_l}
  (L-\ell_{m+1}\openone)\cdots(L-\ell_k\openone)P^{({\rm others})} = 0,
\end{equation}
where $\ell_i$ are the eigenvalues of $L$ on the components $W_i$.
Eq.~(\ref{characteristic_l}) implies a relation of the form
\begin{equation}
\label{reduce_l}
  L^{k-m} = c_1L^{k-m-1} + \cdots + c_{k-m}\openone 
        + c^\prime_1P^{(1)} + \cdots + c^\prime_mP^{(m)},
\end{equation}
with coefficients $c_i$ and $c^\prime_i$. Eq.~(\ref{reduce_l}) is used in the
following to reduce powers of $L$.

In the decomposition~(\ref{decomposition_a2}) for $\su(3)$, the following
projectors have a very simple form:
\begin{equation}
\label{proj_adj_a2}
  P^{[1]}_{jk,pq} = \frac{1}{8}\delta_{jk}\delta_{pq}\,,\qquad
  P^{[8_S]}_{jk,pq} = \frac{3}{5}d_{jkr}d_{pqr}\,,\qquad
  P^{[8_A]}_{jk,pq} = \frac{1}{3}f_{jkr}f_{pqr}\,,
\end{equation}
where $[8_S]$ and $[8_A]$ denote the adjoint representation in the symmetric
and antisymmetric part of the decomposition~(\ref{decomposition_a2}),
respectively.

The eigenvalues of $C_{V\otimes V}$ on the representations $[1]$, $[8]$ and
$[27]$ are $0$, $1$ and $\slantfrac{8}{3}$, respectively, in view of the
normalization defined in eq.~(\ref{casimir_def}). We thus have
$\ell^{[1]}=-1$, $\ell^{[8]}=-\slantfrac{1}{2}$ and
$\ell^{[27]}=\slantfrac{1}{3}$.  Eq.~(\ref{reduce_l}) for the symmetric part
of~(\ref{decomposition_a2}) reads
\begin{equation}
\label{chareq_a2}
  L\openone_S = \frac{1}{3}\openone_S - \frac{4}{3}P^{[1]}
                  -\frac{5}{6}P^{[8_S]},
\end{equation}
where
$(\openone_S)_{jk,pq}=\frac{1}{2}(\delta_{jp}\delta_{kq}+\delta_{jq}\delta_{kp})$
projects onto the symmetric part of the tensor product. In terms of the $f$-
and $d$-tensors of $\su(3)$, eq.~(\ref{chareq_a2}) reads
\begin{equation}
\label{chareq_a2ind}
  f_{jpr}f_{kqr}+f_{jqr}f_{kpr} = -\delta_{jp}\delta_{kq}-\delta_{jq}\delta_{kp} 
    + \delta_{jk}\delta_{pq} + 3\,d_{jkr}d_{pqr}\,.
\end{equation}
This is a well known identity. It coincides eq.~(2.23) in~\mycite{MaSu68}.

A similar argument could be applied to the antisymmetric part of the
decomposition~(\ref{decomposition_a2}), but this would only give the Jacobi
identity as a relation. Reversing the argument, we can use the Jacobi identity
to determine the constituents of the antisymmetric part. If $V_{\rm adj}$
denotes the adjoint representation of $\g$, there is always an adjoint
component in the antisymmetric part of $V_{\rm adj}\otimes V_{\rm adj}$
determined by the projector
\begin{equation}
  P^{[{\rm adj}_A]}_{jk,pq} = C_{jkr}C_{pqr}\,,
\end{equation}
or, in the $\su(3)$ example, by the last entry in~(\ref{proj_adj_a2}). Because
of~(\ref{def_l}), the Jacobi identity is the same as
\begin{mathletters}
\label{antisymmetric}
\begin{equation}
  L\openone_A=-\frac{1}{2}P^{[{\rm adj}_A]},
\end{equation}
where
${(\openone_A)}_{jk,pq}=\frac{1}{2}(\delta_{jp}\delta_{kq}-\delta_{jq}\delta_{kp})$
is the antisymmetrizer. This implies
\begin{equation}
  L^2\openone_A=-\frac{1}{2}L\openone_A.
\end{equation}%
\end{mathletters}%
Hence the minimal equation of $L$ on the antisymmetric subspace, for all $\g$,
takes the form
\begin{equation}
\label{antisymmetric_minimal}
  L(L+\frac{1}{2})\openone_A=0.
\end{equation}
We conclude that the only representations contained in the antisymmetric
subspace have eigenvalues $0$ or $-\slantfrac{1}{2}$ of $L$, \ie\ eigenvalues
$2$ or $1$ of $C_{V\otimes V}$. We employ this fact in
section~\ref{subsect_g2} in the discussion of the exceptional simple Lie
algebra $g_2$ to associate representations of the same dimension with the
symmetric versus antisymmetric part. The relations~(\ref{antisymmetric}) are
furthermore useful to get rid of the symmetrizer in eq.~(\ref{chareq_a2}).

Now we apply a Jacobi type identity twice to eq.~(\ref{chareq_a2ind}) and
derive
\begin{equation}
\label{chareq_a2symm}
  d_{r(jk}d_{p)qr} = \frac{1}{3}\delta_{(jk}\delta_{p)q}\,,
\end{equation}
obtained as eq.~(2.22) in~\mycite{MaSu68} from the characteristic equation of
$A=a_j\lambda_j$, $a_j\in\C$. We see~(\ref{chareq_a2symm}) easily allows us to
calculate
\begin{equation}
  \tr(\lambda_{(i}\lambda_j\lambda_{k)}\lambda_\ell)
  = 2\delta_{(ij}\delta_{k)\ell}\,,
\end{equation}
and 
\begin{equation}
\label{tran_su3}
  \tr A^4=\frac{1}{2}{(\tr A^2)}^2,
\end{equation}
which reflects the fact (trivial in this context) that the fourth order
Casimir operator~\cite{Ok77,Ok82} of $\su(3)$ is not primitive. Our second
method described in section~\ref{subsect_traces} analyzes these dependencies
systematically.

For $\su(4)$, referring to the decomposition~(\ref{decomposition_a3}), we note
that the projectors $P^{[1]}$ and $P^{[15_S]}$ belonging to the symmetric part
have a simple form, but the two remaining ones, $P^{[20]}$ and $P^{[84]}$,
have not. The eigenvalues of $C_{V\otimes V}$ are $0$, $1$, $\slantfrac{3}{2}$
and $\slantfrac{5}{2}$. Therefore $\ell^{[1]}=-1$,
$\ell^{[15]}=-\slantfrac{1}{2}$, $\ell^{[20]}=-\slantfrac{1}{4}$ and
$\ell^{[84]}=\slantfrac{1}{4}$. Since there are two projectors which are not
known immediately, we derive instead of~(\ref{chareq_a2}) an equation for the
square of $L$:
\begin{equation}
\label{chareq_a3}
  L^2\openone_S = \frac{1}{16}\openone_S+\frac{15}{16}P^{[1]}+\frac{3}{16}P^{[15_S]},
\end{equation}
from which the term proportional to $L\openone_S$ accidentally vanishes. This
means
\begin{equation}
\label{chareq_a3inda}
  f_{jmr}f_{knr}(f_{mps}f_{nqs}+f_{mqs}f_{nps})
   = \delta_{jp}\delta_{kq}+\delta_{jq}\delta_{kp} + 2\,\delta_{jk}\delta_{pq} 
   + 2\,d_{jkr}d_{pqr}\,.
\end{equation}
Using the relations~(\ref{antisymmetric}), we obtain
\begin{equation}
  L^2 = \frac{1}{16}\openone_S+\frac{15}{16}P^{[1]}+\frac{3}{16}P^{[15_S]}+\frac{1}{4}P^{[15_A]}
\end{equation}
or
\begin{equation}
\label{chareq_a3ind}
  2\,f_{jmr}f_{knr}f_{mps}f_{nqs} = \delta_{jp}\delta_{kq} + \delta_{jq}\delta_{kp}
    +2\,\delta_{jk}\delta_{pq}+2\,d_{jkr}d_{pqr}+2\,f_{jkr}f_{pqr}\,.
\end{equation}
This is equivalent to eq.~(A.5) in~\mycite{AzMa98} and can be seen, by use of
eq.~(\ref{jacobi_mixed}), to imply eq.~(A.11) in~\mycite{AzMa98} in the case
$n=4$.

The method described above has the advantage that it does not rely on a
well-developed $f$- and $d$-tensor technique but instead refers to the
representation theory of the adjoint representation. It is therefore
applicable even in cases where the tensor calculations are of significant
difficulty (\eg\ $g_2$, section~\ref{subsect_g2}) or not available at all
(\eg\ $f_4$, $e_6$, section~\ref{subsect_f4} and~\ref{subsect_e6}).

As a by-product, it allows us to determine the remaining projectors, and thus
Clebsch-Gordan coefficients, via
\begin{equation}
  P^{(j)}=\prod_{i\neq j}\frac{L-\ell_i\openone}{\ell_j-\ell_i}\,,
\end{equation}
where relation~(\ref{reduce_l}) reduces the power of the operator $L$
appearing in the expression for $P^{(j)}$ to at most $L^{k-m-1}$.

In $\su(3)$ there is, of course~\cite{MaSu68},
\begin{equation}
  P^{[27]}_{jk,pq} = {\Bigl(\openone_S-P^{[1]}-P^{[8_S]}\Bigr)}_{jk,pq} =
    \frac{1}{2}(\delta_{jp}\delta_{kq} + \delta_{jq}\delta_{kp})
    - \frac{1}{8}\delta_{jk}\delta_{pq} - \frac{3}{5}d_{jkr}d_{pqr}\,.
\end{equation}
But in $\su(4)$ we have the less obvious but clearly useful results
\begin{mathletters}
\begin{eqnarray}
  P^{[20]}_{jk,pq} &=& \frac{1}{4}(\delta_{jp}\delta_{kq} + \delta_{jq}\delta_{kp}
     + f_{jpr}f_{kqr} + f_{jqr}f_{kpr}) 
     - \frac{1}{6}\delta_{jk}\delta_{pq}-\frac{1}{2}d_{jkr}d_{pqr}\,,\\
  P^{[84]}_{jk,pq} &=& \frac{1}{4}(\delta_{jp}\delta_{kq} + \delta_{jq}\delta_{kp}
     - f_{jpr}f_{kqr} - f_{jqr}f_{kpr}) 
     + \frac{1}{10}\delta_{jk}\delta_{pq}+\frac{1}{6}d_{jkr}d_{pqr}\,.
\end{eqnarray}%
\end{mathletters}%
Since we use the characteristic (more precisely: minimal) polynomial of $L$
in~(\ref{characteristic_l}), our method fails to separate projectors onto
representations which have the same eigenvalues of the quadratic Casimir
operator as \eg\ two conjugate representations do have. Possible treatments of
this point involve the use of higher order Casimir operators or the explicit
consideration of a conjugation operation in the field of complex numbers
\cite{Su69}.

While we have examined here only the decomposition of the adjoint
representation (in order to derive relations involving the structure
constants), the same method can be applied to other representations to obtain
convenient expressions for some of the projectors occurring there.

\subsection{Relations of trace polynomials}
\label{subsect_traces}

Our second method is based on the characteristic equation of representation
matrices $A=a_jx_j$ of elements $a_jX_j\in\g$, $a_j\in\C$. It enables us to
derive explicit relations between invariant polynomials of $\g$ and to express
non-primitive polynomials explicitly in terms of primitive ones. This gives
rise to further class~2 identities that generalize~(\ref{tran_su3}).

Let $G$ be any connected Lie group with Lie algebra $\g$. Due to a theorem of
Chevalley (see \eg~\mycite{Hu80,Va84}) the algebra $I_\g$ of polynomials on
$\g$ that are invariant under the adjoint action of $G$ is isomorphic to a
polynomial algebra in $\ell=\rk\g$ indeterminates. Since the generalized
Casimir operators~\cite{Ok77,EnKi80,Hu80,Va84} and the centre of the universal
enveloping algebra $U(\g)$ can be constructed from these polynomials, an
explicit description of them is desired.

Let $A=a_jx_j$ be the matrix of an arbitrary element $a_jX_j\in\g$ in the
$d$-dimensional defining representation. One possible strategy~\cite{Mo98} is
to consider the characteristic polynomial of $A$
\begin{equation}
\label{char_general}
  \chi_A(t)=\det(A-t\openone)=\sum_{j=0}^dp_{d-j}(A)\,t^j,
\end{equation}
whose coefficients $p_k(A)$ are homogeneous polynomials of degree $k$ in the
elements of the matrix $A$. They are $G$-invariant by construction. Depending
on the Lie algebra, a certain set of $\ell$ polynomials can be selected that
generates $I_\g$ freely~\cite{Hu80,Va84}. These generators are therefore
called primitive, and their degrees are a property of $\g$ itself. For easy
reference, we include the table of their degrees for the simple Lie algebras
in table~\ref{tab_primitive}.

\begin{table}
\begin{center}
\begin{tabular}{cllc}
\hhline
\hglue 1.6cm & simple Lie algebra $\g$ & degrees & \hglue 1.6cm\\
\hline
&$a_\ell,\ell\geq 1$ & $2,3,\ldots,\ell+1$\\
&$b_\ell,\ell\geq 2$ & $2,4,\ldots,2\ell$\\
&$c_\ell,\ell\geq 3$ & $2,4,\ldots,2\ell$\\
&$d_\ell,\ell\geq 4$ & $2,4,\ldots,2\ell-2,\ell$\\
&$e_6$               & $2,5,6,8,9,12$\\
&$e_7$               & $2,6,8,10,12,14,18$\\
&$e_8$               & $2,8,12,14,18,20,24,30$\\
&$f_4$               & $2,6,8,12$\\
&$g_2$               & $2,6$\\
\hhline
\end{tabular}
\end{center}
\caption{
  \label{tab_primitive}Degrees of primitive invariant polynomials of the
  simple Lie algebras.}
\end{table}

Another way of constructing manifestly $G$-invariant polynomials uses the
trace polynomials $\tr(A^k)$, $k\in\N$, which we study below. Again it is
desired to select a subset of algebraically independent polynomials which have
the required degrees and therefore freely generate $I_\g$.

The fact that eq.~(\ref{char_general}) only gives primitive $G$-invariant
polynomials up to a certain degree $m$ (at least not higher than the dimension
$d$ of the defining representation) was exploited in~\mycite{Mo98} to derive
relations expressing $\tr A^k$, $k>m$, as polynomials in the lower degree
traces. But the method in~\mycite{Mo98} needs additional input in certain
situations. For example, the identities for $g_2$ that account for the
non-primitive nature of the quartic Casimir operator of $g_2$ are not
themselves generated by the method. Thus, we seek a systematic approach which
expresses {\sl all\/} non-primitive $\tr A^j$ in terms of primitive ones.

In order to perform explicit calculations involving the $\tr A^j$, we exploit
their invariance under a change of basis in their representation space in
order to diagonalize $A$. The resulting matrix can be seen to belong to the
Cartan subalgebra $\g_0$ of $\g$ and therefore may be written
\begin{equation}
  A=\gamma_jh_j,\quad 1\leq j\leq\ell=\rk\g,
\end{equation}
where $h_1,\ldots,h_\ell$ span $\g_0$. They are determined by the weights of
the representation. The expressions $\tr A^k$ are polynomials of degree $k$ in
the $\ell$ indeterminates $\gamma_1,\ldots,\gamma_\ell$. To express a
non-primitive $\tr A^k$ in terms of the primitive ones means to write it as a
linear combination of products of the primitive ones. The problem of finding
the relations is reduced to a (very manageable) problem in linear algebra.

It follows from the construction that the resulting relations do not
depend on the choice of basis of $\g_0$ and therefore not on the normalization
or orthogonality of the weights.

Let us first consider the defining representation $[3]$ of $\su(3)$. The
choice of a convenient basis for the diagonal form of
$A=\frac{1}{2}a_j\lambda_j$ leads to
\begin{equation}
\label{csa_a2d}
  A=\alpha\lambda_3+\frac{\beta}{2}(\lambda_3+\sqrt{3}\lambda_8)
   =\diag(\alpha+\beta,-\alpha,-\beta),
\end{equation}
with indeterminates $\alpha,\beta$, therefore
\begin{mathletters}
\begin{eqnarray}
  \tr A   &=& 0,\\
  \tr A^2 &=& 2\,(\alpha^2+\alpha\beta+\beta^2),\\
  \tr A^3 &=& 3\,(\alpha^2\beta+\alpha\beta^2),\\
  \tr A^4 &=& 2\,(\alpha^4+2\alpha^3\beta+3\alpha^2\beta^2+2\alpha\beta^3+\beta^4),\\
  &\ldots&\nn
\end{eqnarray}%
\end{mathletters}%
Because of the degrees of primitive invariant polynomials of $\su(3)$
(table~\ref{tab_primitive}) we can select $\tr A^2$ and $\tr A^3$ as
generators of the algebra of invariant polynomials. The other $\tr A^k$ can be
seen to satisfy the relations
\begin{mathletters}
\label{rel_a2}
\begin{eqnarray}
  \tr A^4 &=& \frac{1}{2}{(\tr A^2)}^2,\\
  \tr A^5 &=& \frac{5}{6}(\tr A^2)(\tr A^3),\\
  \tr A^6 &=& \frac{1}{4}{(\tr A^2)}^3 + \frac{1}{3}{(\tr A^3)}^2,\\
  &\ldots&\nn
\end{eqnarray}%
\end{mathletters}%
and many more (see appendix~\ref{subapp_a2}). It is also easy to compute the
characteristic polynomial of $A$,
\begin{equation}
\label{char_poly_a2d}
  \chi_A(t) = t^3 - \frac{1}{2}(\tr A^2)\,t - \frac{1}{3}(\tr A^3),
\end{equation}
which has been explicitly known for a long time~\cite{MaSu68}. We are also
able to reproduce eq.~(6.5) and~(6.7) from~\mycite{AzMa98} by translating the
relations into the language of generalized $d$-tensors (see~\mycite{Su90}). We
have, for example,
\begin{mathletters}
\label{d_technology}
\begin{eqnarray}
  \tr (\lambda_{(i_1}\lambda_{i_2}\lambda_{i_3)}) &=& 2d_{(i_1i_2i_3)}\,,\\
  \tr (\lambda_{(i_1}\lambda_{i_2}\lambda_{i_3}\lambda_{i_4)})
    &=& \frac{4}{n}\delta_{(i_1i_2}\delta_{i_3i_4)}
        + 2d^{(4)}_{(i_1i_2i_3i_4)}\,,\\
  \tr (\lambda_{(i_1}\cdots\lambda_{i_5)})
    &=& \frac{2}{n}d_{(i_1i_2i_3}\delta_{i_4i_5)}
        + 2d^{(5)}_{(i_1i_2i_3i_4i_5)}\,,\\
  &\ldots&\nn
\end{eqnarray}%
\end{mathletters}%
the first of which defines the third order invariant tensor already
present in $\su(3)$. Whereas in $\su(3)$ the higher order
$d^{(k)}_{(i_1\ldots i_k)}$ can be reduced to this and to the second
order $\delta_{i_1i_2}$ tensor, for general $\su(n)$ there are in
total $n-1$ primitive of them. Using the
expressions~(\ref{d_technology}), our method easily reproduces the
symmetrized forms of eq.~(27) to~(29) in~\mycite{Su90} for $\su(3)$
and likewise for higher $\su(n)$.

A similar procedure works with the matrices $F=a_jF_j$, $a_j\in\C$,
${(F_j)}_{k\ell}=if_{j\ell k}$, of the adjoint representation of $\su(3)$.
Here we have $\tr(F_jF_k)=3\delta_{jk}$.  Using the same basis of the Cartan
subalgebra as in eq.~(\ref{csa_a2d}), we write
\begin{eqnarray}
  F&=&2\alpha F_3+\beta(F_3+\sqrt{3}F_8)\nn\\
   &=&\diag(\alpha+2\beta,\alpha-\beta,2\alpha+\beta,0,0,-2\alpha-\beta,
    -\alpha+\beta,-\alpha-2\beta),
\end{eqnarray}
so that $\tr F^{2k-1}=0$ for all $k\in\N$, and
\begin{mathletters}
\begin{eqnarray}
  \tr F^2 &=& 12(\alpha^2+\alpha\beta+\beta^2),\\
  \tr F^4 &=& 36(\alpha^4+2\alpha^3\beta+3\alpha^2\beta^2+2\alpha\beta^3+\beta^4),\\
  &\ldots&\nn
\end{eqnarray}%
\end{mathletters}%
Although these $\tr F^{2k}$ have unsuitable degrees to generate the algebra of
invariant polynomials freely, their relations are worth noting,
\begin{mathletters}
\label{resf_a2}
\begin{eqnarray}
\label{resf4_a2}
  \tr F^4 &=&   \frac{1}{4}{(\tr F^2)}^2,\\
\label{resf8_a2}
  \tr F^8 &=& - \frac{5}{192}{(\tr F^2)}^4
              + \frac{2}{3}(\tr F^2)(\tr F^6),\\
  &\ldots&\nn
\end{eqnarray}%
\end{mathletters}%
as well as is the characteristic polynomial of $F$,
\begin{equation}
\label{char_poly_a2aa}
\chi_F(t) = t^8 -\frac{1}{2}(\tr F^2)\,t^6 + \frac{1}{16}{(\tr F^2)}^2\,t^4 
            + \Bigl(\frac{1}{96}{(\tr F^2)}^3 - \frac{1}{6}(\tr F^6)\Bigr)\,t^2.
\end{equation}
In eq.~(\ref{resf_a2}) we have listed only those relations which express a
certain $\tr F^{2k}$ in terms of lower degree traces. Since the $\tr F^{2k}$
fail to define a third order invariant, the $\tr F^{2k}$ are unable to
generate the algebra of invariant polynomials freely. It turns out, further,
that in $a_4=\su(5)$ (see appendix~\ref{subapp_a4}) the $\tr F^{2k}$,
$2k\in\{2,4,6,8,10\}$ cannot be written as polynomials in the lower degree
traces. The fact that there are five of them while the rank of the algebra is
only four, does not contradict any theorem because they do not generate the
algebra freely. Instead they should satisfy more complicated (\ie\ higher
order) relations which we have not analyzed systematically.

Eq.~(\ref{resf4_a2}) is already of interest; it yields
\begin{equation}
  \tr(F_{(i}F_jF_kF_{\ell)}) = \frac{9}{4}\delta_{(ij}\delta_{k\ell)}\,,
\end{equation}
which agrees (A.12) of~\mycite{AzMa98}, but has been more simply derived here.
As a further check on our work, we confirmed that our result for $\tr F^8$
agrees with relations derived in a different way in~\mycite{Mo98}.

Since we are using the same basis of the Cartan subalgebra for the diagonal
forms of both $A$ and $F$, we are able to express the $\tr F^{2k}$ in terms of
the primitive elements $\tr A^2$ and $\tr A^3$, namely
\begin{mathletters}
\label{resfn_a2}
\begin{eqnarray}
\label{resfn_a2a}
\tr F^2 &=& 6(\tr A^2),\\
\tr F^4 &=& 9{(\tr A^2)}^2,\\
\label{resfn_a2c}
\tr F^6 &=& \frac{33}{2}{(\tr A^2)}^3 -18{(\tr A^3)}^2,\\
\label{resfn_a2d}
\tr F^8 &=& \frac{129}{4}{(\tr A^2)}^4 -72(\tr A^2){(\tr A^3)}^2,\\
&\ldots&\nn
\end{eqnarray}%
\end{mathletters}%
where eq.~(\ref{resfn_a2a}) reflects the relative normalizations of the
matrices $A$ and $F$. The characteristic polynomial of $F$ can thus be given
in a form more useful than (\ref{char_poly_a2aa})
\begin{equation}
\label{char_poly_a2ab}
\chi_F(t) = t^8 -3(\tr A^2)\,t^6 + \frac{9}{4}{(\tr A^2)}^2\,t^4 
            + \Bigl(-\frac{1}{2}{(\tr A^2)}^3 + 3{(\tr A^3)}^2\Bigr)\,t^2.
\end{equation}
From eq.~(\ref{resfn_a2c}) and eq.~(\ref{resfn_a2d}), for example, we can show
that
\begin{mathletters}
\begin{eqnarray}
\label{resf6_a2}
  \tr(F_{(i_1}\cdots F_{i_6)}) 
    &=& \frac{33}{16}\delta_{(i_1i_2}\delta_{i_3i_4}\delta_{i_5i_6)}
      - \frac{9}{8}d_{(i_1i_2i_3}d_{i_4i_5i_6)}\,,\\
\label{resf8_a2i}
  \tr(F_{(i_1}\cdots F_{i_8)})
    &=& \frac{129}{64}\delta_{(i_1i_2}\delta_{i_3i_4}\delta_{i_5i_6}\delta_{i_7i_8)}
      - \frac{9}{4}\delta_{(i_1i_2}d_{i_3i_4i_5}d_{i_6i_7i_8)}\,,
\end{eqnarray}%
\end{mathletters}%
which are new results, the first being related to but not easily derived from
(A.21) in~\mycite{AzMa98}. The second could be also derived from
eq.~(\ref{resf8_a2}) and eq.~(\ref{resf6_a2}).

Since this method of determining the explicit relations among invariant
polynomials only relies on the properties of the relevant polynomials, it is
well suitable for automatization using computer algebra systems. Applications
to other rank $2$ algebras including $g_2$ can still be easily performed by
hand, whereas for higher rank the polynomials become quite lengthy. We
performed many calculations, including these for low degrees in the rank $2$
examples by hand, and used Maple to confirm our results and to handle the more
complicated computations.

%
\section{Applications and selected results}
%
\label{sect_appl}

In this section we describe the most important results concerning other Lie
algebras than those considered in the examples of section~\ref{sect_method},
and we comment on some aspects of them. A more comprehensive list of results
obtained by our second method is contained in appendix~\ref{app_casimir}.

\subsection{The simple Lie algebras $a_\ell$, $\ell\geq 3$}
\label{subsect_an}

In this paragraph we summarize our results for the simple Lie algebras
$a_\ell$, $\ell\geq 3$ (or $\su(n)$, $n=\ell+1\geq 4$). In order to write them
in a fashion independent of $n$, we characterize the representations by their
highest weight which we specify in terms of the fundamental weights
$\Lambda_1,\ldots,\Lambda_\ell$ in standard form~\cite{Hu80,Co84}. The adjoint
representation is therefore $(1,0,\ldots,0,1)$, and we find the decomposition
into irreducible components
\begin{eqnarray}
\label{decomposition_an}
  && (1,0,\ldots,0,1)\otimes(1,0,\ldots,0,1)\nn\\
  &=& \underbrace{(0,\ldots,0)\oplus(1,0,\ldots,0,1)\oplus(0,1,0,\ldots,0,1,0)
                \oplus(2,0,\ldots,0,2)}_{\rm symmetric}\\
  &\oplus& \underbrace{(1,0,\ldots,0,1)\oplus(2,0,\ldots,0,1,0)
                \oplus(0,1,0,\ldots,0,2)}_{\rm antisymmetric}\nn
\end{eqnarray}
where, for example, $(0,1,0,\ldots,0,1,0)$ corresponds to the highest weight
$\Lambda=\Lambda_2+\Lambda_{\ell-1}$. This reduces to \eg\ $(0,2,0)$ if
$\ell=3$. The dimensions and the eigenvalues of the quadratic Casimir operator
and of $L$ can be directly computed from the Cartan matrix and are listed in
table~\ref{tab_casimir_an}.

\beforetab
\begin{table}
\begintab{lcllcc}
\hhline
$i$ & representation of $a_\ell$ & $d$ & $\Lambda$ & $C$ & $L$\\
\hline
$1$ & $(0,\ldots,0)$         & $1$ 
    & $0$                          & $0$                        & $-1$ \\
$2$ & $(1,0,\ldots,0,1)$     & $\ell(\ell+2)$         
    & $\Lambda_1+\Lambda_\ell$     & $1$                        & $-\frac{1}{2}$ \\
$3$ & $(0,1,0,\ldots,0,1,0)$ & $\frac{1}{4}{(\ell+1)}^2(\ell+2)(\ell-2)$ 
    & $\Lambda_2+\Lambda_{\ell-1}$ & $\frac{2\ell}{\ell+1}$     & $-\frac{1}{\ell+1}$ \\
$4$ & $(2,0,\ldots,0,2)$     & $\frac{1}{4}\ell{(\ell+1)}^2(\ell+4)$ 
    & $2\Lambda_1+2\Lambda_\ell$   & $2\,\frac{\ell+2}{\ell+1}$ & $\frac{1}{\ell+1}$ \\
$5$ & $(2,0,\ldots,0,1,0)$   & $\frac{1}{4}\ell(\ell+2)(\ell+3)(\ell-1)$ 
    & $2\Lambda+\Lambda_{\ell-1}$  & $2$                        & $0$ \\
$6$ & $(0,1,0,\ldots,0,2)$   & $\frac{1}{4}\ell(\ell+2)(\ell+3)(\ell-1)$ 
    & $\Lambda_2+2\Lambda_{\ell}$  & $2$                        & $0$ \\
\hhline
\endtab
\caption{ \label{tab_casimir_an}The irreducible components occurring
  in the tensor product of the adjoint representation of $a_\ell$,
  $\ell\geq 3$, with itself. The representations are of dimension $d$
  and have the highest weight $\Lambda$. The eigenvalues of the
  quadratic Casimir operator $C$ and of $L$ (see eq.~(\ref{def_l}))
  are also given. The number $i$ is used to refer to them in the
  text.}
\end{table}
\aftertab

In the decomposition~(\ref{decomposition_an}), the following projectors
$P^{(i)}$ (where $i$ is used as shown in the first column of
table~\ref{tab_casimir_an}) are initially known:
\begin{mathletters}
\begin{eqnarray}
P^{(1)}_{jk,pq}   &=& \frac{1}{n^2-1}\delta_{jk}\delta_{pq}\,,\\
P^{(2_S)}_{jk,pq} &=& \frac{n}{n^2-4}d_{jkr}d_{pqr}\,,\\
P^{(2_A)}_{jk,pq} &=& \frac{1}{n}f_{jkr}f_{pqr}\,.
\end{eqnarray}%
\end{mathletters}%
Relation~(\ref{reduce_l}) gives in this case
\begin{mathletters}
\begin{eqnarray}
\label{resf4_an}
  L^2 &=& \frac{1}{n^2}\openone_S+\frac{n^2-1}{n^2}P^{(1)}
                    +\frac{n^2-4}{4n^2}P^{(2_S)}-\frac{1}{2}L\openone_A,\\
\label{resf4_anb}
  4f_{jmr}f_{knr}f_{mps}f_{nqs} &=& 2(\delta_{jp}\delta_{kq}+\delta_{jq}\delta_{kp})
    + 4\delta_{jk}\delta_{pq} + n\,(d_{jkr}d_{pqr} + f_{jkr}f_{qpr})\,.
\end{eqnarray}%
\end{mathletters}%
Applying eq.~(\ref{jacobi_mixed}) we see that (\ref{resf4_anb}) coincides
eq.~(A.5) and implies eq.~(A.11) in~\mycite{AzMa98}. The remaining projectors
$P^{(3)}$ and $P^{(4)}$ in the symmetric part can also be found:
\begin{mathletters}
\begin{eqnarray}
P^{(3)}_{jk,pq} &=& \frac{1}{4}(\delta_{jp}\delta_{kq} + \delta_{jq}\delta_{kp})
                 - \frac{1}{2(n-1)}\delta_{jk}\delta_{pq}
                 - \frac{n}{4(n-2)}d_{jkr}d_{pqr}\\
              && + \frac{1}{4}(f_{jpr}f_{kqr} + f_{jqr}f_{kpr})\,,\nn\\
P^{(4)}_{jk,pq} &=& \frac{1}{4}(\delta_{jp}\delta_{kq} + \delta_{jq}\delta_{kp})
                 + \frac{1}{2(n+1)}\delta_{jk}\delta_{pq}
                 + \frac{n}{4(n+2)}d_{jkr}d_{pqr}\\
              && - \frac{1}{4}(f_{jpr}f_{kqr} + f_{jqr}f_{kpr})\,.\nn
\end{eqnarray}%
\end{mathletters}%
The $\su(4)$ example presented in section~\ref{subsect_methodproj} is a
special case of these results. From eq.~(\ref{resf4_anb}) we derive
\begin{equation}
  \tr(F_{(i}F_jF_kF_{\ell)}) = 2\delta_{(ij}\delta_{k\ell)} 
    + \frac{n}{4}d_{r(ij}d_{k\ell)r}
\end{equation}
which implies
\begin{equation}
  \tr F^4 = 6{(\tr A^2)}^2 + 2n(\tr A^4).
\end{equation}
Note that because of our normalization conventions $\tr
(F_jF_k)=f_{pqj}f_{pqk}=n\delta_{jk}$ and thus $\tr F^2 = 2n\,(\tr A^2)$.
From the relations obtained with our second method from eq.~(\ref{restab_a3}),
we obtain \eg\ for $\su(4)$
\begin{equation}
  \tr (F_{(i_1}\cdots F_{i_6)}) 
    = \delta_{(i_1i_2}\delta_{i_3i_4}\delta_{i_5i_6)} 
      + \frac{9}{4}\delta_{(i_1i_2}d^r{}_{i_3i_4}d_{i_5i_6)r}
      - \frac{13}{12}d_{(i_1i_2i_3}d_{i_4i_5i_6)}\,,
\end{equation}
and for $\su(5)$ (\cf\ eq.~(\ref{restab_a4}))
\begin{equation}
  \tr (F_{(i_1}\cdots F_{i_6)}) 
    = \frac{65}{64}\delta_{(i_1i_2}\delta_{i_3i_4}\delta_{i_5i_6)} 
      + \frac{75}{32}\delta_{(i_1i_2}d^r{}_{i_3i_4}d_{i_5i_6)r}
      - \frac{25}{24}d_{(i_1i_2i_3}d_{i_4i_5i_6)}\,.
\end{equation}

\subsection{The simple Lie algebras $b_\ell$, $\ell\geq 2$}
\label{subsect_bn}

We describe the Lie algebras $b_\ell$ (or $\so(n)$, $n=2\ell+1$) by a
basis $(x_j)$ of $n\times n$ antisymmetric hermitean matrices $x_j$ in
the defining representation. They are normalized such that
$\tr(x_jx_k)=2\delta_{jk}$. We define the structure constants
$c_{jk\ell}$ using $[x_i,x_j]=ic_{jk\ell}x_\ell$. Since the
symmetrized product $\{x_j,x_k\}$ is a symmetric matrix, we have
\begin{equation}
\label{multiply_bn}
  \{x_j,x_k\}=\frac{4}{n}\delta_{jk}\openone + d_{jk\alpha}y_\alpha\,,
\end{equation}
where the $y_\alpha$ span the space of traceless symmetric $n\times n$
matrices and have the normalization $\tr(y_\alpha
y_\beta)=2\delta_{\alpha\beta}$. The coefficients $d_{jk\alpha}$ form a
tensor with the properties $d_{jk\alpha}=d_{kj\alpha}$ and
$d_{jj\alpha}=0$. The collection of all $x_i$ and $y_\alpha$ can serve
as a set of Gell-Mann matrices of $\su(n)$.

It is to be noted that here for $b_\ell$ as well as below for $c_\ell$
and for $d_\ell$, we index the basis vectors $x_j$ by a single letter
rather than by the (perhaps more usual) method that uses an index pair,
as in $M_{ab}=E_{ab}-E_{ba}=-M_{ba}$, where
${(E_{ab})}_{cd}=\delta_{ac}\delta_{bd}$. This is first suitably
convenient for all our purposes and second allows a uniform
treatment of all Lie algebras.

Since the Cartan-Killing form reads in our basis $\tr (\ad x_j\circ \ad
x_k)=2(2\ell-1)\delta_{jk}$, the structure constants $c_{jkr}$ are
related to the $C_{jkr}$ used in our general discussion by
$C_{jkr}=-\slantfrac{c_{jkr}}{\sqrt{2(2\ell-1)}}$.

\beforetab
\begin{table}
\begintab{llllcc}
\hhline
$i$ & representation of $b_\ell$ & $d$ & $\Lambda$ & $C$
  & $L$\\
\hline
$1$ & $(0,\ldots,0)$         & $1$             
  & $0$                   & $0$                        & $-1$ \\
$2$ & $(0,1,0,\ldots,0)$     & $\ell(2\ell+1)$ 
  & $\Lambda_2$           & $1$                        & $-\frac{1}{2}$ \\
$3$ & $(0,2,0,\ldots,0)$     & $\frac{1}{3}(\ell-1)(\ell+1)(2\ell+1)(2\ell+3)$ 
  & $2\Lambda_2$          & $\frac{4\ell}{2\ell-1}$    & $\frac{1}{2\ell-1}$ \\
$4$ & $(0,0,0,1,0,\ldots,0)$ & $\frac{1}{6}\ell(\ell-1)(2\ell-1)(2\ell+1)$
  & $\Lambda_4$           & $\frac{2(2\ell-3)}{2\ell-1}$ & $-\frac{2}{2\ell-1}$ \\
$5$ & $(2,0,\ldots,0)$       & $\ell(2\ell+3)$
  & $2\Lambda_1$          & $\frac{2\ell+1}{2\ell-1}$  & $-\frac{2\ell-3}{2(2\ell-1)}$ \\
$6$ & $(1,0,1,0,\ldots,0)$   & $\frac{1}{2}\ell(\ell-1)(2\ell+1)(2\ell+3)$
  & $\Lambda_1+\Lambda_3$ & $2$                        & $0$ \\
\hhline
\endtab
\caption{ \label{tab_casimir_bn}The irreducible components occurring
  in the tensor product of the adjoint representation of $b_\ell$,
  $\ell\geq 5$, with itself. Here $i$, $d$, $\Lambda$, $C$ and $L$ are
  as for table~\ref{tab_casimir_an}. The formulae for dimensions and
  Casimir eigenvalues are also valid for the corresponding
  representations in the cases $\ell=2,3,4$. We comment on this fact
  in the text.}
\end{table}
\aftertab

Table~\ref{tab_casimir_bn} lists the representations which are
relevant to the decomposition of the tensor product of the adjoint
representation with itself for the cases $\ell\geq 5$
\begin{mathletters}
\begin{eqnarray}
  && (0,1,0,\ldots,0)\otimes (0,1,0,\ldots,0)\nn\\
  &=& \underbrace{(0,\ldots,0)\oplus(2,0,\ldots,0)
        \oplus(0,0,0,1,0,\ldots,0)\oplus(0,2,0,\ldots,0)}_{\rm symmetric}\\
  &\oplus& \underbrace{(0,1,0,\ldots,0)
        \oplus (1,0,1,0,\ldots,0)}_{\rm antisymmetric}.\nn
\end{eqnarray}
The highest weights written in terms of the fundamental weights have
special forms for $\ell=2,3,4$. The above decomposition reads in these
cases
\begin{eqnarray}
  (0,2)\otimes(0,2) &=& \underbrace{(0,0)\oplus (2,0)\oplus
                        (1,0)\oplus (0,4)}_{\rm symmetric}\\
                    & & \oplus\underbrace{(0,2)
                        \oplus (1,2)}_{\rm antisymmetric},\nn\\
  (0,1,0)\otimes(0,1,0) &=& \underbrace{(0,0,0)\oplus (2,0,0)\oplus
                            (0,0,2)\oplus (0,2,0)}_{\rm symmetric}\\
                        & & \oplus\underbrace{(0,1,0)
                            \oplus (1,0,2)}_{\rm antisymmetric},\nn\\
  (0,1,0,0)\otimes(0,1,0,0) &=& \underbrace{(0,0,0,0)\oplus (2,0,0,0)\oplus
                                (0,0,0,2)\oplus (0,2,0,0)}_{\rm symmetric}\\
                            & & \oplus\underbrace{(0,1,0,0)
                                \oplus (1,0,1,0)}_{\rm antisymmetric}.\nn
\end{eqnarray}%
\end{mathletters}%
But nevertheless, the formulae in table~\ref{tab_casimir_bn} for
dimensions and Casimir eigenvalues are valid for arbitrary $\ell\geq
2$.

The following projectors $P^{(i)}$ (where $i$ refers to the rows of
table~\ref{tab_casimir_bn}) are easily written down:
\begin{mathletters}
\label{proj_bn}
\begin{eqnarray}
P^{(1)}_{jk,pq} &=& \frac{2}{n(n-1)}\delta_{jk}\delta_{pq}\,,\\
P^{(5)}_{jk,pq} &=& \frac{1}{2(n-2)}d_{jk\alpha}d_{pq\alpha}\,,\\
P^{(2)}_{jk,pq} &=& \frac{1}{2(n-2)}c_{jkr}c_{pqr}\,.
\end{eqnarray}%
\end{mathletters}%
The projector $P^{(5)}$ has been constructed using the tensor
$d_{jk\alpha}$ from eq.~(\ref{multiply_bn}). A careful analysis of the
tensors involved in relations like~(\ref{multiply_bn}) shows that this
is in fact a projector onto an irreducible component of the symmetric
tensor product of the adjoint representation. A consideration of the
relevant dimensions furthermore allows us to identify the representation
to be $(2,0,\ldots,0)$.

Relation~(\ref{reduce_l}) gives in this case
\begin{mathletters}
\label{resbl}
\begin{eqnarray}
  L^2 &=& \frac{2}{{(n-2)}^2}\openone_S + \frac{(n-1)(n-4)}{{(n-2)}^2}P^{(1)}
          +\frac{n-8}{4(n-2)}P^{(5)}\\
      & &   -\frac{1}{n-2}L\openone_S-\frac{1}{2}L\openone_A,\nn\\
  c_{jmr}c_{knr}c_{mps}c_{nqs} &=& 4\,(\delta_{jp}\delta_{kq} + \delta_{jq}\delta_{kp})
       +\frac{8(n-4)}{n}\delta_{jk}\delta_{pq} + \frac{n-8}{2}d_{jk\alpha}d_{pq\alpha}\\
    & &+\frac{n}{2}c_{jpr}c_{kqr}-\frac{n-4}{2}c_{jqr}c_{kpr}\,,\nn
\end{eqnarray}%
\end{mathletters}%
which implies
\begin{equation}
\tr (F_{(i_1}F_{i_2}F_{i_3}F_{i_4)}) = \frac{16(n-2)}{n}\delta_{(i_1i_2}\delta_{i_3i_4)}
   +\frac{n-8}{2}d_{(i_1i_2}{}^\alpha d_{i_3i_4)\alpha}\,.
\end{equation}
Since we have
\begin{equation}
\tr (x_{(i_1}x_{i_2}x_{i_3}x_{i_4)}) = \frac{4}{n}\delta_{(i_1i_2}\delta_{i_3i_4)}
  + \frac{1}{2}d_{(i_1i_2}{}^\alpha d_{i_3i_4)\alpha}\,,
\end{equation}
we finally derive
\begin{equation}
\label{resf4_bn}
\tr F^4 = 3{(\tr A^2)}^2 + (n-8)(\tr A^4)
\end{equation}
in agreement with the formulae listed in appendix~\ref{subapp_bn}
which have been derived by our second method. The matrices $F_j$
are normalized in such a way that $\tr
(F_jF_k)=c_{pqj}c_{pqk}=2(n-2)\delta_{jk}$ and thus $\tr F^2=(n-2)\tr
A^2$. The simplest of the relations from appendix~\ref{subapp_b2} for
$b_2=\so(5)$ imply, for example,
\begin{equation}
  \tr(x_{(i_1}\cdots x_{i_6)}) =
    \frac{1}{5}\delta_{(i_1i_2}\delta_{i_3i_4}\delta_{i_5i_6)}
    + \frac{3}{4}\delta_{(i_1i_2}d_{i_3i_4}{}^\alpha d_{i_5i_6)\alpha}
\end{equation}
and
\begin{equation}
  \tr(F_{(i_1}\cdots F_{i_6)}) =
    \frac{93}{5}\delta_{(i_1i_2}\delta_{i_3i_4}\delta_{i_5i_6)}
    - \frac{21}{4}\delta_{(i_1i_2}d_{i_3i_4}{}^\alpha d_{i_5i_6)\alpha}\,.
\end{equation}
The remaining projectors in the symmetric part can be found from eq.~(\ref{resbl}):
\begin{mathletters}
\label{projoth_bn}
\begin{eqnarray}
 P^{(3)}_{jk,pq} &=& \frac{1}{3}(\delta_{jp}\delta_{kq}+\delta_{jq}\delta_{kp})
   + \frac{2(n-4)}{3n(n-2)}\delta_{jk}\delta_{pq}
   + \frac{n-8}{12(n-2)}d_{jk\alpha}d_{pq\alpha}\\
  & & -\frac{1}{12}(c_{jpr}c_{kqr}+c_{jqr}c_{kpr}),\nn\\
 P^{(4)}_{jk,pq} &=& \frac{1}{6}(\delta_{jp}\delta_{kq}+\delta_{jq}\delta_{kp})
   - \frac{2}{3n}\delta_{jk}\delta_{pq}
   - \frac{1}{12}d_{jk\alpha}d_{pq\alpha}\\
  & & +\frac{1}{12}(c_{jpr}c_{kqr}+c_{jqr}c_{kpr}).\nn
\end{eqnarray}%
\end{mathletters}%

\subsection{The simple Lie algebras $c_\ell$, $\ell\geq 2$}
\label{subsect_cn}

We describe the Lie algebras $c_\ell$ (or $\sp(2n)$, $n=\ell$) by a
basis $(x_j)$ of $2n\times 2n$ traceless hermitean matrices $x_j$
in the defining representation satisfying 
\begin{equation}
  Jx_jJ^{-1}=-x_j^T,\qquad J=\pmatrix{0&\openone_n\cr -\openone_n&0}.
\end{equation}
They are normalized in such a way that $\tr(x_jx_k)=2\delta_{jk}$. We write
the structure constants as $[x_i,x_j]=ic_{jk\ell}x_\ell$. Given these $x_j$,
the matrices $Jx_j$ span the space of symmetric $2n\times 2n$ matrices
(including the pure trace). If we extend the set of the $x_i$ to a basis of
$\su(2n)$ by adding further basis vectors $y_\alpha$, the $Jy_\alpha$ are
antisymmetric, and the $y_\alpha$ thus satisfy
\begin{equation}
\label{antisymplectic}
  Jy_\alpha J^{-1}=y_\alpha^T.
\end{equation}
Since the symmetrized product $\{x_j,x_k\}$ also
satisfies~(\ref{antisymplectic}), we have
\begin{equation}
\label{multiply_cn}
  \{x_j,x_k\} = \frac{2}{n}\delta_{jk}\openone + d_{jk\alpha}y_\alpha\,,
\end{equation}
because the $y_\alpha$ together with the unit matrix span the space of
solutions of eq.~(\ref{antisymplectic}). The coefficients
$d_{jk\alpha}$ occurring here form a tensor which satisfies
$d_{jk\alpha}=d_{kj\alpha}$ and $d_{jj\alpha}=0$.

Since the Cartan-Killing form reads in our basis $\tr (\ad x_j\circ \ad
x_k)=4(\ell+1)\delta_{jk}$, the structure constants $c_{jkr}$ are
related to the $C_{jkr}$ used in our general discussion by
$C_{jkr}=-\slantfrac{c_{jkr}}{(2\sqrt{\ell+1})}$.

\beforetab
\begin{table}
\begintab{llllcc}
\hhline
$i$ & representation of $c_\ell$ & $d$ & $\Lambda$ & $C$
  & $L$ \\
\hline
$1$ & $(0,\ldots,0)$     & $1$             
  & $0$                    & $0$                      & $-1$ \\
$2$ & $(2,0,\ldots,0)$   & $\ell(2\ell+1)$ 
  & $2\Lambda_1$           & $1$                      & $-\frac{1}{2}$ \\
$3$ & $(4,0,\ldots,0)$   & $\frac{1}{6}\ell(\ell+1)(2\ell+1)(2\ell+3)$
  & $4\Lambda_1$           & $\frac{2(\ell+2)}{\ell+1}$ & $\frac{1}{\ell+1}$ \\
$4$ & $(0,2,0,\ldots,0)$ & $\frac{1}{3}\ell(\ell-1)(2\ell-1)(2\ell+3)$
  & $2\Lambda_2$           & $\frac{2\ell+1}{\ell+1}$ & $-\frac{1}{2(\ell+1)}$ \\
$5$ & $(0,1,0,\ldots,0)$ & $(\ell-1)(2\ell+1)$
  & $\Lambda_2$            & $\frac{\ell}{\ell+1}$    & $-\frac{\ell+2}{2(\ell+1)}$ \\
$6$ & $(2,1,0,\ldots,0)$ & $\frac{1}{2}\ell(\ell-1)(2\ell+1)(2\ell+3)$
  & $2\Lambda_1+\Lambda_2$ & $2$                      & $0$ \\
\hhline
\endtab
\caption{ \label{tab_casimir_cn}The irreducible components occurring
  in the tensor product of the adjoint representation of $c_\ell$,
  $\ell\geq 2$, with itself. Here $i$, $d$, $\Lambda$, $C$ and $L$ are
  as for table~\ref{tab_casimir_an}.}
\end{table}
\aftertab

Table~\ref{tab_casimir_cn} lists the representations which are
relevant to the decomposition of the tensor product of the adjoint
representation with itself
\begin{eqnarray}
  && (2,0,\ldots,0)\otimes (2,0,\ldots,0)\nn\\
  &=& \underbrace{(0,\ldots,0)\oplus(0,1,0,\ldots,0)
        \oplus(4,0,\ldots,0)\oplus(0,2,0,\ldots,0)}_{\rm symmetric}\\
  &\oplus& \underbrace{(2,0,\ldots,0)
        \oplus (2,1,0,\ldots,0)}_{\rm antisymmetric}.\nn
\end{eqnarray}

The following projectors $P^{(i)}$ (where $i$ refers to the rows of
table~\ref{tab_casimir_cn}) are easily written down:
\begin{mathletters}
\begin{eqnarray}
P^{(1)}_{jk,pq} &=& \frac{1}{n(2n+1)}\delta_{jk}\delta_{pq}\,,\\
P^{(5)}_{jk,pq} &=& \frac{1}{4(n+1)}d_{jk\alpha}d_{pq\alpha}\,,\\
P^{(2)}_{jk,pq} &=& \frac{1}{4(n+1)}c_{jkr}c_{pqr}\,,
\end{eqnarray}%
\end{mathletters}%
where $P^{(5)}$ has been constructed using the tensor $d_{jk\alpha}$
from eq.~(\ref{multiply_cn}).

Relation~(\ref{reduce_l}) reads in this case
\begin{mathletters}
\begin{eqnarray}
  L^2 &=& \frac{1}{2{(n+1)}^2}\openone_S 
          + \frac{(2n+1)(n+2)}{2{(n+1)}^2}P^{(1)}
          + \frac{n+4}{4(n+1)}P^{(5)}\\
      & & + \frac{1}{2(n+1)}L\openone_S - \frac{1}{2}L\openone_A,\nn\\
  c_{jmr}c_{knr}c_{mps}c_{nqs} 
      &=& 4\,(\delta_{jp}\delta_{kq}+\delta_{jq}\delta_{kp})
          + \frac{8(n+2)}{n}\delta_{jk}\delta_{pq}
          + (n+4)\,d_{jk\alpha}d_{pq\alpha}\\
      & & + n\,c_{jpr}c_{kqr}-(n+2)\,c_{jqr}c_{kpr}\,,\nn
\end{eqnarray}%
\end{mathletters}%
which implies
\begin{eqnarray}
  \tr(F_{(i_1}F_{i_2}F_{i_3}F_{i_4}) 
    = \frac{16(n+1)}{n}\delta_{(i_1i_2}\delta_{i_3i_4)} 
      + (n+4)\,d_{(i_1i_2}{}^\alpha d_{i_3i_4)\alpha}
\end{eqnarray}
and
\begin{equation}
  \tr F^4 = 3{(\tr A^2)}^2 + 2(n+4)(\tr A^4).
\end{equation}
This is consistent with the results presented in
appendix~\ref{subapp_cn}. The matrices $F_j$ are normalized in such a
way that $\tr(F_jF_k)=c_{pqj}c_{pqk}=4(n+1)\delta_{jk}$ and thus $\tr
F^2=2(n+1)\tr A^2$. The remaining projectors in the symmetric part
are
\begin{mathletters}
\begin{eqnarray}
  P^{(3)}_{jk,pq} &=& \frac{1}{6}(\delta_{jp}\delta_{kq}+\delta_{jq}\delta_{kp})
    + \frac{1}{3n}\delta_{jk}\delta_{pq}+\frac{1}{12}d_{jk\alpha}d_{pq\alpha}\\
  & & - \frac{1}{12}(c_{jpr}c_{kqr}+c_{jqr}c_{kpr}),\nn\\
  P^{(4)}_{jk,pq} &=& \frac{1}{3}(\delta_{jp}\delta_{kq}+\delta_{jq}\delta_{kp})
    - \frac{2(n+2)}{3n(2n+1)}\delta_{jk}\delta_{pq}
    - \frac{n+4}{12(n+1)}d_{jk\alpha}d_{pq\alpha}\\
  & & + \frac{1}{12}(c_{jpr}c_{kqr}+c_{jqr}c_{kpr}).\nn
\end{eqnarray}%
\end{mathletters}%

\subsection{The simple Lie algebras $d_\ell$, $\ell\geq 3$}

In order to describe the Lie algebras $d_\ell$, $\ell\geq 3$ (or
$\so(n)$, $n=2\ell$), we choose the same basis as for odd $n$ in
section~\ref{subsect_bn}. Because of the different relation between
$\ell$ and $n$, we find $\tr (\ad x_j\circ \ad
x_k)=4(\ell-1)\delta_{jk}$ and therefore
$C_{jkr}=-\slantfrac{c_{jkr}}{(2\sqrt{\ell-1})}$. 

\beforetab
\begin{table}
\begintab{llllcc}
\hhline
$i$ & representation of $d_\ell$ & $d$ & $\Lambda$ & $C$
  & $L$\\
\hline
$1$ & $(0,\ldots,0)$         & $1$              
  & $0$                   & $0$                      & $-1$ \\
$2$ & $(0,1,0,\ldots,0)$     & $\ell(2\ell-1)$
  & $\Lambda_2$           & $1$                      & $-\frac{1}{2}$ \\
$3$ & $(0,2,0,\ldots,0)$     & $\frac{1}{3}\ell(\ell+1)(2\ell-3)(2\ell+1)$
  & $2\Lambda_2$          & $\frac{2\ell-1}{\ell-1}$ & $\frac{1}{2(\ell-1)}$ \\
$4$ & $(0,0,0,1,0,\ldots,0)$ & $\frac{1}{6}\ell(\ell-1)(2\ell-3)(2\ell-1)$
  & $\Lambda_4$           & $\frac{2(\ell-2)}{\ell-1}$ & $-\frac{1}{\ell-1}$ \\
$5$ & $(2,0,\ldots,0)$       & $(\ell+1)(2\ell-1)$
  & $2\Lambda_1$          & $\frac{\ell}{\ell-1}$    & $-\frac{\ell-2}{2(\ell-1)}$ \\
$6$ & $(1,0,1,0,\ldots,0)$   & $\frac{1}{2}\ell(\ell+1)(2\ell-3)(2\ell-1)$
  & $\Lambda_1+\Lambda_3$ & $2$                      & $0$ \\
\hhline
\endtab
\caption{ \label{tab_casimir_dn}The irreducible components occurring
  in the tensor product of the adjoint representation of $d_\ell$,
  $\ell\geq 6$, with itself. We comment on the special cases
  $\ell=3,4,5$ in the text.}
\end{table}
\aftertab

Table~\ref{tab_casimir_dn} lists the representations which are
relevant to the decomposition of the tensor product of the adjoint
representation with itself in the cases $\ell\geq 6$:
\begin{mathletters}
\begin{eqnarray}
\label{decomposition_dn}
  && (0,1,0,\ldots,0)\otimes (0,1,0,\ldots,0)\nn\\
  &=& \underbrace{(0,\ldots,0)\oplus(2,0,\ldots,0)
        \oplus(0,0,0,1,0,\ldots,0)\oplus(0,2,0,\ldots,0)}_{\rm symmetric}\\
  &\oplus& \underbrace{(0,1,0,\ldots,0)
        \oplus (1,0,1,0,\ldots,0)}_{\rm antisymmetric}.\nn
\end{eqnarray}
The special cases for $\ell=3,4,5$ are
\begin{eqnarray}
\label{decompose_d3}
  (0,1,1)&\otimes&(0,1,1)\nn\\
  &=& \underbrace{(0,0,0)\oplus(2,0,0)\oplus
                            (0,1,1)\oplus(0,2,2)}_{\rm symmetric}\\
                        & & \oplus\underbrace{(0,1,1)\oplus
                            \biggl((1,0,2)\oplus(1,2,0)\biggr)}_{\rm antisymmetric},\nn\\
\label{decompose_d4}
  (0,1,0,0)&\otimes&(0,1,0,0)\nn\\
  &=& \underbrace{(0,0,0,0)\oplus(2,0,0,0)\oplus\biggl((0,0,2,0)\oplus(0,0,0,2)\biggr)
        \oplus(0,2,0,0)}_{\rm symmetric}\\
  & & \oplus\underbrace{(0,1,0,0)\oplus(1,0,1,1)}_{\rm antisymmetric},\nn\\
  (0,1,0,0,0)&\otimes&(0,1,0,0,0)\nn\\
  &=& \underbrace{(0,0,0,0,0)\oplus(2,0,0,0,0)\oplus(0,0,0,1,1)
        \oplus(0,2,0,0,0)}_{\rm symmetric}\\
  & & \oplus\underbrace{(0,1,0,0,0)\oplus(1,0,1,0,0)}_{\rm antisymmetric}.\nn
\end{eqnarray}%
\end{mathletters}%
The structure of the decomposition is somewhat exceptional for
$\ell=3$ (because $d_3$ is isomorphic with $a_3$) and for $\ell=4$
(because of the higher symmetry of the Dynkin diagram of $d_4$). We
have indicated by additional parentheses, in (\ref{decompose_d3}) and
(\ref{decompose_d4}), the cases in which a representation corresponding
to one piece in~(\ref{decomposition_dn}) has decomposed into even
smaller constituents. Nevertheless, the formulae in
table~\ref{tab_casimir_dn} for dimensions and Casimir eigenvalues are
useful for arbitrary $\ell\geq 3$. In the case of further
decomposition, the table lists the sum of the dimensions, and both
constituents turn out to have the same Casimir eigenvalue.

If the projectors $P^{(i)}$ and the Casimir eigenvalues $C$ from
table~\ref{tab_casimir_dn} are written using $n=2\ell$, they have
exactly the same form as for the algebras $b_\ell$. Therefore we have
the same results as eqs.~(\ref{proj_bn}) to (\ref{resf4_bn})
and~(\ref{projoth_bn}).

\subsection{The exceptional simple Lie algebra $g_2$}
\label{subsect_g2}

For the structure and the construction of representations of $g_2$ see, for
example, \mycite{GuTz56,BeDr62,Ra76,Ok95}. We use for the defining
representation of $g_2$ a suitable set of $14$ traceless hermitean $7\times 7$
matrices $x_j$ which have the additional properties
\begin{equation}
  x_j^T=-x_j,\quad\tr(x_jx_k)=2\delta_{jk}\,,
\end{equation}
and write the structure constants $c_{jk\ell}$ such that
$[x_j,x_k]=i\,c_{jk\ell}x_k$. Note that they are related to the
constants $C_{jk\ell}$ from the general discussion by
$C_{jk\ell}=-\slantfrac{c_{jk\ell}}{\sqrt{8}}$.  The space of $7\times
7$ traceless hermitean matrices (the matrices of the defining
representation of $a_6=\su(7)$) involves $21$ antisymmetric matrices
which span the $b_3=\so(7)$ subalgebra of $a_6$, and $28$ symmetric
matrices $y_\alpha$, $1\leq\alpha\leq 28$. We do not need to introduce
here the $7$ antisymmetric matrices $z_a$, $1\leq a\leq 7$, which lie
outside the $g_2$ subalgebra of $b_3$, but we do need the
$y_\alpha$. They satisfy
\begin{equation}
  y_\alpha^T=y_\alpha,\quad 
    \tr(y_\alpha y_\beta)=2\delta_{\alpha\beta},\quad
  \tr(x_j y_\alpha)=0,
\end{equation}
and are related to the $x_j$ by
\begin{equation}
\label{g2_product}
  x_jx_k = \frac{2}{7}\delta_{jk} + \frac{1}{2}\,ic_{jk\ell}x_\ell 
         + \frac{1}{2}d_{jk\alpha}y_\alpha,
\end{equation}
where $d_{jk\alpha}=d_{kj\alpha}$ and $d_{jj\alpha}=0$. Of course, complete
control of $g_2$ technology depends on consideration of $x_i$, $y_\alpha$ and
$z_a$, and the various isotropic tensors that enter product laws like
eq.~(\ref{g2_product}). A full treatment of these matters will be presented
elsewhere~\cite{Ma99}. Here we quote class~1 results as needed and attend
to our main purpose, that of deriving class~2 results. It is worth noting that
the set of all $x_i$, $z_a$, $y_\alpha$ can be viewed as a set of $48$
Gell-Mann type matrices $\lambda_A$ of $a_6=\su(7)$.

\beforetab
\begin{table}
\begintab{lrlcc}
\hhline
 representation of $g_2$ &$d$          & $\Lambda$    & $C$ & $L$\\
\hline
 $(0,0)$        &$1$          & $0$          & $0$              & $-1$ \\
 $(0,1)$        &$7$          & $\Lambda_2$  & $\frac{1}{2}$    & $-\frac{3}{6}$ \\
 $(1,0)$        &$14$         & $\Lambda_1$  & $1$              & $-\frac{1}{2}$ \\
 $(0,2)$        &$27$         & $2\Lambda_2$ & $\frac{7}{6}$    & $-\frac{5}{12}$ \\
 $(2,0)$        &$77$         & $2\Lambda_1$ & $\frac{5}{2}$    & $\frac{1}{4}$ \\
 $(0,3)$        &$77^\prime$  & $3\Lambda_2$ & $2$              & $0$ \\
\hhline
\endtab
\caption{
  \label{tab_casimir_g2}Irreducible representations of $g_2$.}
\end{table}
\aftertab

Table~\ref{tab_casimir_g2} contains information about the irreducible
representations of $g_2$ relevant to the application of our first
method. $[7]$ is the defining representation, $[14]$ the adjoint. The
tensor product of the adjoint representation with itself decomposes into
irreducible components as follows
\begin{equation}
  [14]\otimes[14] = \underbrace{[1]\oplus[27]\oplus[77]}_{\rm symmetric}
     \oplus \underbrace{[14]\oplus[77^\prime]}_{\rm antisymmetric}.
\end{equation}
In order to decide which of the $[77]$ or $[77^\prime]$ representations occur
in the symmetric versus antisymmetric part, we consider their
$C_{V\otimes V}$ eigenvalues. Only $[77^\prime]$ gives the eigenvalue
$2$ and therefore, according to the discussion of
eq.~(\ref{antisymmetric_minimal}), belongs to the antisymmetric part.

The following projectors are initially known:
\begin{mathletters}
\begin{eqnarray}
  P^{[1]}_{jk,pq}  &=& \frac{1}{14}\delta_{jk}\delta_{pq}\,,\\
\label{proj_g2b}
  P^{[27]}_{jk,pq} &=& \frac{9}{32}d_{jk\alpha}d_{pq\alpha}\,,\\
  P^{[14]}_{jk,pq} &=& \frac{1}{8}\,c_{jkr}c_{pqr}\,,
\end{eqnarray}%
\end{mathletters}%
in virtue of the identities
\begin{mathletters}
\begin{eqnarray}
\label{didentity_g2}
c_{jkp}c_{jkq}          &=& 8\delta_{pq}\,,\\
d_{jk\alpha}d_{jk\beta} &=& \frac{32}{9}\delta_{\alpha\beta}\,.
\end{eqnarray}%
\end{mathletters}%
As for $\su(3)$ in section~\ref{subsect_methodproj} we deal with the
characteristic equation and, noting that in our basis we have $L_{jk,pq} =
-\frac{1}{8}c_{jpr}c_{kqr}$, derive
\begin{mathletters}
\label{chareq_g2}
\begin{eqnarray}
  L\openone_S &=& \frac{1}{4}\openone_S - \frac{5}{4}P^{[1]}
    - \frac{2}{3}P^{[27]},\\
\label{chareq_g2b}
  c_{jpr}c_{kqr} + c_{jqr}c_{kpr} &=& - 2(\delta_{jp}\delta_{kq}
    + \delta_{jq}\delta_{kp}) + \frac{10}{7}\delta_{jk}\delta_{pq} 
    + 3d_{jk\alpha}d_{pq\alpha}\,.
\end{eqnarray}%
\end{mathletters}%
By two applications of the Jacobi identity, (\ref{chareq_g2b}) yields
\begin{equation}
\label{result_g2}
  d_{(jk}{}^\alpha d_{pq)\alpha} = \frac{6}{7}\delta_{(jk}\delta_{pq)}\,.
\end{equation}
The simplification process used these class~1 identities
\begin{mathletters}
\begin{eqnarray}
  c_{pjq}c_{qkr}c_{r\ell p}           &=& -4c_{jk\ell}\,,\\
  d_{jk\alpha}d_{\ell m\alpha}c_{kmp} &=& \frac{20}{7}c_{j\ell p}\,,\\
  c_{pjq}c_{qkr}d_{jk\alpha} &=& \frac{10}{3}d_{pr\alpha}\,.                     
\end{eqnarray}%
\end{mathletters}%
The results~(\ref{chareq_g2}) and~(\ref{result_g2}) are new, as is
their convenient and therefore important method of derivation. We note
the check that~(\ref{result_g2})
implies~(\ref{didentity_g2}). From~(\ref{result_g2}), we can obtain
\begin{equation}
  \tr(x_{(i}x_jx_kx_{\ell)}) = \delta_{(ij}\delta_{k\ell)}\,,
\end{equation}
and hence if $A=a_jx_j$, $a_j\in\C$, the important result
\begin{equation}
  \tr A^4=\frac{1}{4}{(\tr A^2)}^2\,,
\end{equation}
which was quoted by Okubo~\cite{Ok77} (without what he termed its
rather involved proof).

If we had to perform our calculations in $g_2$ without an explicit
form of the projector $P^{[27]}$ (which we do indeed know thanks to
the $d$-tensor calculus available), the procedure would have been more
like the treatment of the $\su(4)$ example given in
section~\ref{subsect_methodproj}, and would have yielded the weaker
results
\begin{mathletters}
\begin{eqnarray}
  L^2 &=& \frac{5}{48}\openone_S + \frac{35}{48}P^{[1]}
          -\frac{1}{6}L\openone_S - \frac{1}{2}L\openone_A,\\
\label{chareq_g2b2}
  c_{jmr}c_{knr}c_{mps}c_{nqs} 
      &=& \frac{10}{3}(\delta_{jp}\delta_{kq} + \delta_{jq}\delta_{kp} + \delta_{jk}\delta_{pq})
          +\frac{8}{3}c_{jpr}c_{kqr} - \frac{4}{3}c_{jqr}c_{kpr}\,.
\end{eqnarray}%
\end{mathletters}%
But with this information, we can still construct the projector in
question, getting
\begin{equation}
  P^{[27]}_{jk,pq} = \frac{3}{16}(\delta_{jp}\delta_{kq} + \delta_{jq}\delta_{kp})
             - \frac{15}{112}\delta_{jk}\delta_{pq}
             + \frac{3}{32}(c_{jpr}c_{kqr} + c_{jqr}c_{kpr})\,,
\end{equation}
which, of course, agrees eq.~(\ref{proj_g2b}) upon use of
(\ref{chareq_g2b}). Also, from eq.~(\ref{chareq_g2b2}) we can derive
\begin{equation}
  \tr (F_{(i}F_jF_kF_{\ell)}) = 10\delta_{(ij}\delta_{k\ell)}\,,
\end{equation}
where ${(F_j)}_{k\ell} = i\,c_{j\ell k}$ are the matrices in the
adjoint representation. Since $\tr (F_jF_k)=8\delta_{jk}$,
this leads to
\begin{equation}
  \tr F^4 = \frac{5}{32}{(\tr F^2)}^2 = \frac{5}{2}{(\tr A^2)}^2.
\end{equation}

This is consistent with the relations found more easily using our
second method, and listed in appendix~\ref{subapp_g2} for $g_2$. From
those relations, we obtain further identities, \eg\
\begin{mathletters}
\begin{eqnarray}
  \tr(x_{(i_1}\cdots x_{i_8)})
    &=& -\frac{5}{192}\delta_{(i_1i_2}\delta_{i_3i_4}\delta_{i_5i_6}\delta_{i_7i_8)}
        +\frac{2}{3}\delta_{(i_1i_2}d^{(6)}_{i_3\cdots i_8)}\,,\\
  \tr(x_{(i_1}\cdots x_{i_{10})})
    &=& -\frac{1}{64}\delta_{(i_1i_2}\delta_{i_3i_4}\delta_{i_5i_6}\delta_{i_7i_8}\delta_{i_9i_{10})}
        +\frac{5}{16}\delta_{(i_1i_2}\delta_{i_3i_4}d^{(6)}_{i_5\cdots i_{10})}\,,
\end{eqnarray}%
\end{mathletters}%
where 
\begin{equation}
  d^{(6)}_{(i_1\cdots i_6)}=\tr(x_{(i_1}\cdots x_{i_6)})
\end{equation}
denotes the sixth order invariant of $g_2$. In particular we observe
that our identity for $\tr A^8$ is consistent with the results
obtained in~\mycite{Mo98}. Furthermore we can also reduce the
symmetric traces in the adjoint representation:
\begin{mathletters}
\begin{eqnarray}
  \tr(F_{(i_1}\cdots F_{i_6)})
    &=& \frac{15}{4}\delta_{(i_1i_2}\delta_{i_3i_4}\delta_{i_5i_6)}
        - 26\,d^{(6)}_{(i_1\cdots i_6)}\,,\\
  \tr(F_{(i_1}\cdots F_{i_8)})
    &=& \frac{515}{96}\delta_{(i_1i_2}\delta_{i_3i_4}\delta_{i_5i_6}\delta_{i_7i_8)}
        - \frac{160}{3}\delta_{(i_1i_2}d^{(6)}_{i_3\cdots i_8)}\,.
\end{eqnarray}%
\end{mathletters}%
Our discussion of $f_4$ in the next section is similar to that of
$g_2$ without referring to explicit $d$-tensor formulae.

\subsection{The exceptional simple Lie algebra $f_4$}
\label{subsect_f4}

\beforetab
\begin{table}
\begintab{lrlcc}
\hhline
 representation of $f_4$ &$d$ & $\Lambda$    & $C$ & $L$\\
\hline
 $(0,0,0,0)$    &$1$    & $0$          & $0$              & $-1$ \\
 $(0,0,0,1)$    &$26$   & $\Lambda_4$  & $\frac{2}{3}$    & $-\frac{2}{3}$ \\
 $(1,0,0,0)$    &$52$   & $\Lambda_1$  & $1$              & $-\frac{1}{2}$ \\
 $(0,0,0,2)$    &$324$  & $2\Lambda_4$ & $\frac{13}{9}$   & $-\frac{5}{18}$ \\
 $(2,0,0,0)$    &$1053$ & $2\Lambda_1$ & $\frac{20}{9}$   & $\frac{1}{9}$ \\
 $(0,1,0,0)$    &$1274$ & $\Lambda_2$  & $2$              & $0$ \\
\hhline
\endtab
\caption{
  \label{tab_casimir_f4}Irreducible representations of $f_4$.}
\end{table}
\aftertab

For the analysis of $f_4$, we use the structure constants $C_{jk\ell}$
in the general notation defined in the beginning of
section~\ref{sect_sun} because no analogue of the $f$- and $d$-tensor
calculus is known. Table~\ref{tab_casimir_f4} contains information
about the irreducible representations of $f_4$ relevant to the
application of our first method.  $[26]$ is the defining
representation, $[52]$ the adjoint. The tensor product of the adjoint
representation with itself decomposes into irreducible components as
follows
\begin{equation}
\label{decomposition_f4}
[52]\otimes[52] = \underbrace{[1]\oplus[324]\oplus[1053]}_{\rm symmetric}
    \oplus \underbrace{[52]\oplus[1274]}_{\rm antisymmetric}.
\end{equation}
We have the relation
\begin{mathletters}
\begin{eqnarray}
  L^2 &=& \frac{5}{162}\openone_S  + \frac{65}{81}P^{[1]}
          - \frac{1}{6}L\openone_S -\frac{1}{2}L\openone_A,\\
\label{resf4b}
  C_{jmr}C_{knr}C_{mps}C_{nqs} 
    &=& \frac{5}{324}(\delta_{jp}\delta_{kq} + \delta_{jq}\delta_{kp}
        + \delta_{jk}\delta_{pq})\\
    & & + \frac{1}{3}C_{jpr}C_{kqr} - \frac{1}{6}C_{jqr}C_{kpr}\,.\nn
\end{eqnarray}%
\end{mathletters}%
and the projectors
\begin{mathletters}
\begin{eqnarray}
  P^{[1]}_{jk,pq}  &=& \frac{1}{52}\delta_{jk}\delta_{pq}\,,\\
  P^{[52]}_{jk,pq} &=& C_{jkr}C_{pqr}\,,\\
  P^{[324]}_{jk,pq}  &=& \frac{1}{7}(\delta_{jp}\delta_{kq} + \delta_{jq}\delta_{kp})
                        -\frac{5}{91}\delta_{jk}\delta_{pq}
                        +\frac{9}{7}(C_{jpr}C_{kqr} + C_{jqr}C_{kpr})\,,\\
  P^{[1053]}_{jk,pq} &=& \frac{5}{14}(\delta_{jp}\delta_{kq} + \delta_{jq}\delta_{kp})
                        + \frac{1}{28}\delta_{jk}\delta_{pq}
                        -\frac{9}{7}(C_{jpr}C_{kqr} + C_{jqr}C_{kpr})\,,\\
  P^{[1274]}_{jk,pq} &=& \frac{1}{2}(\delta_{jp}\delta_{kq} - \delta_{jq}\delta_{kp})
                        - C_{jkr}C_{pqr}.
\end{eqnarray}%
\end{mathletters}%
Furthermore we derive from~(\ref{resf4b})
\begin{equation}
  \tr(F_{(i}F_jF_kF_{\ell)}) = \frac{5}{108}\delta_{(ij}\delta_{k\ell)}\,,
\end{equation}
and therefore
\begin{equation}
  \tr F^4 = \frac{5}{108}{(\tr F^2)}^2,
\end{equation}
where ${(F_j)}_{k\ell} = C_{j\ell k}$ are the matrices of the adjoint
representation, which obey $\tr (F_jF_k) = -\delta_{jk}$.

Since there is no generally accepted definition of the invariant tensors of $f_4$, we use
$\tr A^k$, $A=a_jx_j$, to define them in a totally symmetric form
\begin{equation}
  d^{(k)}_{i_1\ldots i_k}:=\tr(x_{(i_1}\cdots x_{i_k)}),\qquad k\in\{2,6,8,12\}.
\end{equation}
Here the $x_j$ are the matrices of the defining representation of $f_4$ and
$d^{(2)}_{i_1i_2}\sim\delta_{i_1i_2}$. The relations that express the
non-primitivity of trace polynomials (see appendix~\ref{subapp_f4}) thus read
\begin{mathletters}
\begin{eqnarray}
  \tr(x_{(i_1}\cdots x_{i_4)}) &=& \frac{1}{2}d^{(2)}_{(i_1i_2}d^{(2)}_{i_3i_4)}\,,\\
  \tr(x_{(i_1}\cdots x_{i_{10})}) &=& 
    \frac{7}{41472}d^{(2)}_{(i_1i_2}d^{(2)}_{i_3i_4}\cdots d^{(2)}_{i_9i_{10})}
    - \frac{7}{144}d^{(2)}_{(i_1i_2}d^{(2)}_{i_3i_4}d^{(6)}_{i_5\ldots i_{10})}\\
  && + \frac{3}{8}d^{(2)}_{(i_1i_2}d^{(8)}_{i_3\ldots i_{10})}\,,
\end{eqnarray}%
\end{mathletters}%
and so on. Furthermore we derive for the matrices ${(F_j)}_{k\ell}$ of the
adjoint representation (in a suitable normalization)
\begin{mathletters}
\begin{eqnarray}
  \tr(F_{(i_1}\cdots F_{i_4)}) &=& \frac{5}{12}d^{(2)}_{(i_1i_2}d^{(2)}_{i_3i_4)}\,,\\
  \tr(F_{(i_1}\cdots F_{i_6)}) &=&
    \frac{5}{36}d^{(2)}_{(i_1i_2}d^{(2)}_{i_3i_4}d^{(2)}_{i_5i_6)}
      - 7d^{(6)}_{(i_1i_2i_3i_4i_5i_6)}\,.
\end{eqnarray}%
\end{mathletters}%

\subsection{The exceptional simple Lie algebra $e_6$}
\label{subsect_e6}

\beforetab
\begin{table}
\begintab{lrlcc}
\hhline
 representation of $e_6$ &$d$ & $\Lambda$             & $C$ & $L$\\
\hline
 $(0,0,0,0,0,0)$ &$1$         & $0$                   & $0$              & $-1$ \\
 $(1,0,0,0,0,0)$ &$27$        & $\Lambda_1$           & $\frac{13}{18}$  & $-\frac{23}{36}$ \\
 $(0,0,0,0,1,0)$ &$27^\prime$ & $\Lambda_5$           & $\frac{13}{18}$  & $-\frac{23}{36}$ \\
 $(0,0,0,0,0,1)$ &$78$        & $\Lambda_6$           & $1$              & $-\frac{1}{2}$ \\
 $(1,0,0,0,1,0)$ &$650$       & $\Lambda_1+\Lambda_5$ & $\frac{3}{2}$    & $-\frac{1}{4}$ \\
 $(0,0,0,0,0,2)$ &$2430$      & $2\Lambda_6$          & $\frac{13}{6}$   & $\frac{1}{12}$ \\
 $(0,0,1,0,0,0)$ &$2925$      & $\Lambda_3$           & $2$              & $0$ \\
\hhline
\endtab
\caption{
  \label{tab_casimir_e6}Irreducible representations of $e_6$.}
\end{table}
\aftertab

In $e_6$ we use again the structure constants $C_{jk\ell}$ in the
general notation. The structure of the decomposition of the adjoint
representation and therefore of the results is similar to what we
found for $f_4$. Table~\ref{tab_casimir_e6} contains information about
the relevant representations. Either $[27]$ or $[27^\prime]$ can play
the role of a defining representation, $[78]$ is the adjoint. The
decomposition is
\begin{equation}
\label{decomposition_e6}
[78]\otimes[78] = \underbrace{[1]\oplus[650]\oplus[2430]}_{\rm symmetric}
    \oplus \underbrace{[78]\oplus[2925]}_{\rm antisymmetric}.
\end{equation}
As before we obtain the relation
\begin{mathletters}
\begin{eqnarray}
L^2 &=& \frac{1}{48}\openone_S + \frac{13}{16}P^{[1]} 
        - \frac{1}{6}L\openone_S - \frac{1}{2}L\openone_A,\\
\label{res_f4}
C_{jmr}C_{knr}C_{mps}C_{nqs} 
    &=& \frac{1}{96}(\delta_{jp}\delta_{kq}+\delta_{jq}\delta_{kp} 
        +\delta_{jk}\delta_{pq})\\
    & &  + \frac{1}{3}C_{jpr}C_{kqr} - \frac{1}{6}C_{jpr}C_{kqr}\,.\nn
\end{eqnarray}%
\end{mathletters}%
The relevant projectors are
\begin{mathletters}
\begin{eqnarray}
  P^{[1]}_{jk,pq}    &=& \frac{1}{78}\delta_{jk}\delta_{pq}\,,\\
  P^{[78]}_{jk,pq}   &=& C_{jkr}C_{pqr}\,,\\
  P^{[650]}_{jk,pq}  &=& \frac{1}{8}(\delta_{jp}\delta_{kq} + \delta_{jq}\delta_{kp})
                        - \frac{1}{24}\delta_{jk}\delta_{pq}
                        + \frac{3}{2}(C_{jpr}C_{kqr} + C_{jqr}C_{kpr})\,,\\
  P^{[2430]}_{jk,pq} &=& \frac{3}{8}(\delta_{jp}\delta_{kq} + \delta_{jq}\delta_{kp})
                        - \frac{3}{104}\delta_{jk}\delta_{pq}
                        - \frac{3}{2}(C_{jpr}C_{kqr} + C_{jqr}C_{kpr})\,,\\
  P^{[2925]}_{jk,pq} &=& \frac{1}{2}(\delta_{jp}\delta_{kq} - \delta_{jq}\delta_{kp})
                        - C_{jkr}C_{pqr}\,.
\end{eqnarray}%
\end{mathletters}%
From the relation~(\ref{res_f4}) we derive
\begin{equation}
  \tr(F_{(i}F_jF_kF_{\ell)}) = \frac{1}{32}\delta_{(ij}\delta_{k\ell)}\,,
\end{equation}
which implies
\begin{equation}
  \tr F^4 = \frac{1}{32}{(\tr F^2)}^2,
\end{equation}
where we have defined the matrices of the adjoint representation as
${(F_j)}_{k\ell} = C_{j\ell k}$, \ie\ $\tr (F_jF_k) = -\delta_{jk}$.

Again there is no common choice for the invariant tensors of $e_6$, so we
define 
\begin{equation}
  d^{(k)}_{i_1\ldots i_k}:=\tr(x_{(i_1}\cdots x_{i_k)}),\qquad k\in\{2,5,6,8,9,12\}.
\end{equation}
Here the $x_j$ are the matrices of the defining representation of $e_6$ and
$d^{(2)}_{i_1i_2}\sim\delta_{i_1i_2}$. The simplest non-trivial relations from
appendix~\ref{subapp_e6} read
\begin{mathletters}
\begin{eqnarray}
  \tr(x_{(i_1}\cdots x_{x_4)}) &=& \frac{1}{12}d^{(2)}_{(i_1i_2}d^{(2)}_{i_3i_4)}\,,\\
  \tr(x_{(i_1}\cdots x_{x_7)}) &=& \frac{7}{24}d^{(2)}_{(i_1i_2}d^{(5)}_{i_3i_4i_5i_6i_7)}\,,
\end{eqnarray}%
\end{mathletters}%
as well as
\begin{mathletters}
\begin{eqnarray}
  \tr(F_{(i_1}\cdots F_{i_4)}) &=& \frac{1}{2}d^{(2)}_{(i_1i_2}d^{(2)}_{i_3i_4)}\,,\\
  \tr(F_{(i_1}\cdots F_{i_6)}) &=& \frac{5}{36}d^{(2)}_{(i_1i_2}d^{(2)}_{i_3i_4}d^{(2)}_{i_5i_6)}
    -6d^{(6)}_{(i_1i_2i_3i_4i_5i_6)}\,,
\end{eqnarray}%
\end{mathletters}%
where ${(F_j)}_{k\ell}$ are the matrices of the adjoint representation (in a
suitable normalization).

\acknowledgements

The research reported in this paper is supported in part by a PPARC grant.
H.~Pfeiffer would like to thank DAAD for a scholarship
``Doktorandenstipendium im Rahmen des gemeinsamen
Hochschulsonderprogramms~III von Bund und L\"andern''. Thanks also are
due to J.~A.~de Azc\'arraga, A.~J.~Mountain and A.~Sudbery for useful
discussions. 

\appendix
%
\section{Appendix: Relations of trace polynomials}
%
\label{app_casimir}

In the appendix we list class~2 relations, specific to each $\g$,
obtained by the method described in section~\ref{subsect_traces} in a
systematic way. If required, our algorithm is able to deal with even
higher order traces.

\subsection{The simple Lie algebras $a_\ell$}

We define the matrices of the defining representations of the Lie algebras
$a_\ell$ by the Gell-Mann matrices $A=\frac{1}{2}a_j\lambda_j$, so that $\tr
A=0$. The adjoint representations are given by ${(F_j)}_{k\ell}=i\,f_{j\ell k}$ 
(where $[\lambda_j,\lambda_k]=2i\,f_{jk\ell}\lambda_\ell$). The odd traces
$\tr F^{2k-1}$, $k\in\N$, vanish. 

\subsubsection{The Lie algebra $a_2$}
\label{subapp_a2}

\paragraph{The defining representation:}

In the defining representation $(1,0)[3]$, the polynomials $\tr A^2$ and $\tr
A^3$ can be taken as generators of the algebra of invariant polynomials. Some
of the results have been given in section~\ref{subsect_traces}. We obviously
do not repeat them. Thus, we have that the $\tr A^k$ satisfy three relations,
given as eq.~(\ref{rel_a2}) for $k=4,5,6$, and also
\begin{mathletters}
\begin{eqnarray}
\tr A^7    &=&   \frac{7}{12}{(\tr A^2)}^2(\tr A^3),\\
\tr A^8    &=&   \frac{1}{8}{(\tr A^2)}^4 + \frac{4}{9}(\tr A^2){(\tr A^3)}^2,\\
\tr A^9    &=&   \frac{3}{8}{(\tr A^2)}^3(\tr A^3) + \frac{1}{9}{(\tr A^3)}^3,\\
\tr A^{10} &=&   \frac{1}{16}{(\tr A^2)}^5 + \frac{5}{12}{(\tr A^2)}^2{(\tr A^3)}^2.
\end{eqnarray}%
\end{mathletters}%
The characteristic polynomial of the matrix $A$ is given by eq.~(\ref{char_poly_a2d}).

\paragraph{The adjoint representation:}

The traces $\tr F^k$, $k\in\N$, in the adjoint representation $(1,1)[8]$ can
be expressed in terms of the primitive polynomials $\tr A^2$ and $\tr A^3$ via
eq.~(\ref{resfn_a2}) for $2k=2,4,6,8$, and also
\begin{mathletters}
\label{tracea2_fa}
\begin{eqnarray}
\tr F^{10} &=&   \frac{513}{8}{(\tr A^2)}^5 
               - \frac{405}{2}{(\tr A^2)}^2{(\tr A^3)}^2,\\
\tr F^{12} &=&   \frac{2049}{16}{(\tr A^2)}^6 
               - 504{(\tr A^2)}^3{(\tr A^3)}^2
               + 54{(\tr A^3)}^4.
\end{eqnarray}%
\end{mathletters}%
The simplest relations of the $\tr F^{2k}$ are given by
eq.~(\ref{resf_a2}) for $2k=4,8$, and also
\begin{mathletters}
\label{tracea2_ff}
\begin{eqnarray}
\tr F^{10} &=& - \frac{1}{64}{(\tr F^2)}^5
               + \frac{5}{16}{(\tr F^2)}^2(\tr F^6),\\
\tr F^{12} &=& - \frac{19}{3072}{(\tr F^2)}^6
               + \frac{5}{48}{(\tr F^2)}^3(\tr F^6)
               + \frac{1}{6}{(\tr F^6)}^2.
\end{eqnarray}%
\end{mathletters}%
The characteristic polynomial of $F$ has been given above in
eqs.~(\ref{char_poly_a2aa}) and~(\ref{char_poly_a2ab}).

\subsubsection{The Lie algebra $a_3$}

\paragraph{The defining representation:}

In the defining representation $(1,0,0)[4]$ of $a_3$, the polynomials $\tr
A^2$, $\tr A^3$ and $\tr A^4$ can be taken as generators of the algebra of
invariant polynomials. The $\tr A^k$ satisfy these relations:
\begin{mathletters}
\begin{eqnarray}
\tr A^5    &=&   \frac{5}{6}(\tr A^2)(\tr A^3),\\
\tr A^6    &=& - \frac{1}{8}{(\tr A^2)}^3
               + \frac{1}{3}{(\tr A^3)}^2
               + \frac{3}{4}(\tr A^2)(\tr A^4),\\
\tr A^7    &=&   \frac{7}{24}{(\tr A^2)}^2(\tr A^3)
               + \frac{7}{12}(\tr A^3)(\tr A^4),\\
\tr A^8    &=& - \frac{1}{16}{(\tr A^2)}^4
               + \frac{4}{9}(\tr A^2){(\tr A^3)}^2
               + \frac{1}{4}{(\tr A^2)}^2(\tr A^4)
               + \frac{1}{4}{(\tr A^4)}^2,\\
\tr A^9    &=&   \frac{1}{9}{(\tr A^3)}^3
               + \frac{3}{4}(\tr A^2)(\tr A^3)(\tr A^4),\\
\tr A^{10} &=& - \frac{1}{64}{(\tr A^2)}^5 
               + \frac{5}{18}{(\tr A^2)}^2{(\tr A^3)}^2
               + \frac{5}{18}{(\tr A^3)}^2(\tr A^4)\\
           & & + \frac{5}{16}(\tr A^2){(\tr A^4)}^2.\nn
\end{eqnarray}%
\end{mathletters}%
The characteristic polynomial of the matrix $A$ is
\begin{equation}
\chi_A(t) = t^4 - \frac{1}{2}(\tr A^2)\,t^2 - \frac{1}{3}(\tr A^3)\,t 
  + \Bigl(\frac{1}{8}{(\tr A^2)}^2 - \frac{1}{4}(\tr A^4)\Bigr).
\end{equation}

\paragraph{The adjoint representation:}

The traces $\tr F^k$, $k\in\N$, in the adjoint representation $(1,0,1)[15]$
can be expressed in terms of the primitive polynomials via
\begin{mathletters}
\label{restab_a3}
\begin{eqnarray}
\tr F^2    &=&   8(\tr A^2),\\
\tr F^4    &=&   6{(\tr A^2)}^2 + 8(\tr A^4),\\
\tr F^6    &=& - {(\tr A^2)}^3 - \frac{52}{3}{(\tr A^3)}^2 
               + 36 (\tr A^2)(\tr A^4),\\
\tr F^8    &=& - \frac{15}{2}{(\tr A^2)}^4
               - \frac{640}{9}(\tr A^2){(\tr A^3)}^2
               + 44{(\tr A^2)}^2(\tr A^4)
               + 72{(\tr A^4)}^2,\\
\tr F^{10} &=& - \frac{23}{4}{(\tr A^2)}^5
               - \frac{1825}{9}{(\tr A^2)}^2{(\tr A^3)}^2
               - 30{(\tr A^2)}^3(\tr A^4)\\
           & & + \frac{20}{9}{(\tr A^3)}^2(\tr A^4)
               + 340(\tr A^2){(\tr A^4)}^2.\nn
\end{eqnarray}%
\end{mathletters}%
The simplest relations of the $\tr F^{2k}$ are the following.
\begin{mathletters}
\begin{eqnarray}
\tr F^8    &=&   \frac{35}{1248}{(\tr F^2)}^4
               - \frac{43}{104}{(\tr F^2)}^2(\tr F^4)
               + \frac{9}{8}{(\tr F^4)}^2
               + \frac{20}{39}(\tr F^2)(\tr F^6),\\
\tr F^{10} &=&   \frac{41}{2496}{(\tr F^2)}^5
               - \frac{295}{1248}{(\tr F^2)}^3(\tr F^4)
               + \frac{35}{52}(\tr F^2){(\tr F^4)}^2\\
           & & + \frac{115}{624}{(\tr F^2)}^2(\tr F^6)
               - \frac{5}{312}(\tr F^4)(\tr F^6).\nn
\end{eqnarray}%
\end{mathletters}%

\subsubsection{The Lie algebra $a_4$}
\label{subapp_a4}

\paragraph{The defining representation:}

In the defining representation $(1,0,0,0)[5]$, the polynomials $\tr A^k$,
$k\in\{2,3,4,5\}$ can be taken as generators of the algebra of invariant
polynomials. The $\tr A^k$ satisfy these relations:
\begin{mathletters}
\begin{eqnarray}
\tr A^6    &=& - \frac{1}{8}{(\tr A^2)}^3
               + \frac{1}{3}{(\tr A^3)}^2
               + \frac{3}{4}(\tr A^2)(\tr A^4),\\
\tr A^7    &=& - \frac{7}{24}{(\tr A^2)}^2(\tr A^3)
               + \frac{7}{12}(\tr A^3)(\tr A^4)
               + \frac{7}{10}(\tr A^2)(\tr A^5),\\
\tr A^8    &=& - \frac{1}{16}{(\tr A^2)}^4
               + \frac{1}{4}{(\tr A^2)}^2(\tr A^4)
               + \frac{1}{4}{(\tr A^4)}^2
               + \frac{8}{15}(\tr A^3)(\tr A^5).
\end{eqnarray}%
\end{mathletters}%
The characteristic polynomial of the matrix $A$ is
\begin{eqnarray}
\chi_A(t) &=& t^5 - \frac{1}{2}(\tr A^2)\,t^3 - \frac{1}{3}(\tr A^3)\,t^2 
        + \Bigl(\frac{1}{8}{(\tr A^2)}^2 - \frac{1}{4}(\tr
        A^4)\Bigr)\,t\\
        & & + \Bigl(\frac{1}{6}(\tr A^2)(\tr A^3)-\frac{1}{5}(\tr A^5)\Bigr).\nn
\end{eqnarray}

\paragraph{The adjoint representation:}

The traces in the adjoint representation $(1,0,0,1)[24]$ of $a_4$ can be
expressed in terms of the primitive polynomials via
\begin{mathletters}
\label{restab_a4}
\begin{eqnarray}
\tr F^2    &=&   10(\tr A^2),\\
\tr F^4    &=&   6{(\tr A^2)}^2 +10(\tr A^4),\\
\tr F^6    &=& - \frac{5}{4}{(\tr A^2)}^3 - \frac{50}{3}{(\tr A^3)}^2
               + \frac{75}{2}(\tr A^2)(\tr A^4),\\
\tr F^8    &=& - \frac{61}{8}{(\tr A^2)}^4
               + \frac{56}{3}(\tr A^2){(\tr A^3)}^2
               + \frac{89}{2}{(\tr A^2)}^2(\tr A^4)\\
           & & + \frac{145}{2}{(\tr A^4)}^2
               - \frac{320}{3}(\tr A^3)(\tr A^5).\nn
\end{eqnarray}%
\end{mathletters}%
There are no relations expressing $\tr F^k$, $k\in\{2,4,6,8,10\}$ as
polynomials of the lower degree ones. The first such relation involves
$\tr F^{12}$. We discussed this fact in section~\ref{subsect_traces}.
\begin{eqnarray}
\tr F^{12} &=&   \frac{13799}{61440000}{(\tr F^2)}^6
               - \frac{8193}{1024000}{(\tr F^2)}^4(\tr F^4)\\
           & & + \frac{3873}{51200}{(\tr F^2)}^2{(\tr F^4)}^2
               - \frac{1957}{12800}{(\tr F^4)}^3
               + \frac{6293}{240000}{(\tr F^2)}^3(\tr F^6)\nn\\
           & & - \frac{1113}{4000}(\tr F^2)(\tr F^4)(\tr F^6)
               + \frac{731}{3750}{(\tr F^6)}^2
               - \frac{759}{6400}{(\tr F^2)}^2(\tr F^8)\nn\\
           & & + \frac{177}{320}(\tr F^4)(\tr F^8)
               + \frac{54}{125}(\tr F^2)(\tr F^{10}).\nn
\end{eqnarray}

\subsection{The simple Lie algebras $b_\ell$}
\label{subapp_bn}

We define the matrices of the defining representations of the Lie algebras
$b_\ell$ by the matrices $A=a_jx_j$ given in section~\ref{subsect_bn}. 
The adjoint representations are defined by ${(F_j)}_{k\ell}=i\,c_{j\ell k}$
(where $[x_i,x_j]=ic_{jk\ell}x_\ell$). The odd traces $\tr A^{2k-1}$ and $\tr
F^{2k-1}$, $k\in\N$, vanish.  

\subsubsection{The Lie algebra $b_2$}
\label{subapp_b2}

\paragraph{The defining representation:}

In the defining representation $(1,0)[5]$ of $b_2$, the polynomials $\tr A^2$
and $\tr A^4$ can be used as generators of the algebra of invariant
polynomials. The others can be expressed in terms of the primitive ones as
follows:
\begin{mathletters}
\begin{eqnarray}
\tr A^6    &=& - \frac{1}{8}{(\tr A^2)}^3 
               + \frac{3}{4}(\tr A^2)(\tr A^4),\\
\tr A^8    &=& - \frac{1}{16}{(\tr A^2)}^4
               + \frac{1}{4}{(\tr A^2)}^2(\tr A^4)
               + \frac{1}{4}{(\tr A^4)}^2,\\
\tr A^{10} &=& - \frac{1}{64}{(\tr A^2)}^5
               + \frac{5}{16}(\tr A^2){(\tr A^4)}^2,\\
\tr A^{12} &=& - \frac{3}{64}{(\tr A^2)}^4(\tr A^4)
               + \frac{3}{16}{(\tr A^2)}^2{(\tr A^4)}^2
               + \frac{1}{16}{(\tr A^4)}^3.
\end{eqnarray}%
\end{mathletters}%
The characteristic polynomial of the matrix $A$ is
\begin{equation}
\chi(t) = t^5 - \frac{1}{2}(\tr A^2)\,t^3 
        + \Bigl(\frac{1}{8}{(\tr A^2)}^2 - \frac{1}{4}(\tr
        A^4)\Bigr)\,t.
\end{equation} 

\paragraph{The adjoint representation:}

The traces in the adjoint representation $(0,2)[10]$ can be expressed in terms
of the primitive polynomials via
\begin{mathletters}
\begin{eqnarray}
\tr F^2    &=&   3(\tr A^2),\\
\tr F^4    &=&   3{(\tr A^2)}^2 - 3(\tr A^4),\\
\tr F^6    &=&   \frac{27}{8}{(\tr A^2)}^3
               - \frac{21}{4}(\tr A^2)(\tr A^4),\\
\tr F^8    &=&   \frac{67}{16}{(\tr A^2)}^4
               - \frac{39}{4}{(\tr A^2)}^2(\tr A^4)
               + \frac{17}{4}{(\tr A^4)}^2,\\
\tr F^{10} &=&   \frac{327}{64}{(\tr A^2)}^5
               - 15{(\tr A^2)}^3(\tr A^4)
               + \frac{165}{16}(\tr A^2){(\tr A^4)}^2,\\
\tr F^{12} &=&   \frac{99}{16}{(\tr A^2)}^6
               - \frac{1395}{64}{(\tr A^2)}^4(\tr A^4)
               + \frac{339}{16}{(\tr A^2)}^2{(\tr A^4)}^2\\
           & & - \frac{63}{16}{(\tr A^4)}^3.\nn
\end{eqnarray}%
\end{mathletters}%
The simplest relations involving the $\tr F^{2k}$ alone are these:
\begin{mathletters}
\begin{eqnarray}
\tr F^6    &=& - \frac{5}{72}{(\tr F^2)}^3
               + \frac{7}{12}(\tr F^2)(\tr F^4),\\
\tr F^8    &=& - \frac{7}{432}{(\tr F^2)}^4
               + \frac{5}{108}{(\tr F^2)}^2(\tr F^4)
               + \frac{17}{36}{(\tr F^4)}^2,\\
\tr F^{10} &=&   \frac{1}{576}{(\tr F^2)}^5
               - \frac{5}{72}{(\tr F^2)}^3(\tr F^4)
               + \frac{55}{144}(\tr F^2){(\tr F^4)}^2,\\
\tr F^{12} &=&   \frac{35}{15552}{(\tr F^2)}^6
               - \frac{187}{5184}{(\tr F^2)}^4(\tr F^4)
               + \frac{25}{216}{(\tr F^2)}^2{(\tr F^4)}^2\\
           & & + \frac{7}{48}{(\tr F^4)}^3.\nn
\end{eqnarray}%
\end{mathletters}%

\subsubsection{The Lie algebra $b_3$}

\paragraph{The defining representation:}

In the defining representation $(1,0,0)[7]$, the polynomials $\tr A^2$, $\tr
A^4$ and $\tr A^6$ can be used as generators of the algebra of invariant
polynomials. The others can be expressed in terms of the primitive ones as
follows:
\begin{mathletters}
\begin{eqnarray}
\tr A^8    &=&   \frac{1}{48}{(\tr A^2)}^4
               - \frac{1}{4}{(\tr A^2)}^2(\tr A^4)
               + \frac{1}{4}{(\tr A^4)}^2
               + \frac{2}{3}(\tr A^2)(\tr A^6),\\
\tr A^{10} &=&   \frac{1}{96}{(\tr A^2)}^5
               - \frac{5}{48}{(\tr A^2)}^3(\tr A^4)
               + \frac{5}{24}{(\tr A^2)}^2(\tr A^6)\\
           & & + \frac{5}{12}(\tr A^4)(\tr A^6),\nn\\
\tr A^{12} &=&   \frac{1}{384}{(\tr A^2)}^6
               - \frac{1}{64}{(\tr A^2)}^4(\tr A^4)
               - \frac{3}{32}{(\tr A^2)}^2{(\tr A^4)}^2\\
           & & + \frac{1}{16}{(\tr A^4)}^3
               + \frac{1}{24}{(\tr A^2)}^3(\tr A^6)
               + \frac{1}{4}(\tr A^2)(\tr A^4)(\tr A^6)\nn\\
           & & + \frac{1}{6}{(\tr A^6)}^2.\nn
\end{eqnarray}%
\end{mathletters}%
The characteristic polynomial of the matrix $A$ is
\begin{eqnarray}
\chi_A(t) &=& t^7 - \frac{1}{2}(\tr A^2)\,t^5 
            + \Bigl( \frac{1}{8}{(\tr A^2)}^2 
              - \frac{1}{4}(\tr A^4)\Bigr)\,t^3\\
        & & + \Bigl(-\frac{1}{48}{(\tr A^2)}^3 
              + \frac{1}{8}(\tr A^2)(\tr A^4)
              - \frac{1}{6}(\tr A^6)\Bigr)\,t.\nn
\end{eqnarray}

\paragraph{The adjoint representation:}

The traces in the adjoint representation $(0,1,0)[21]$ can be expressed in
terms of the primitive polynomials via
\begin{mathletters}
\begin{eqnarray}
\tr F^2    &=&   5(\tr A^2),\\
\tr F^4    &=&   3{(\tr A^2)}^2 - (\tr A^4),\\
\tr F^6    &=&   15(\tr A^2)(\tr A^4) - 25(\tr A^6),\\
\tr F^8    &=& - \frac{121}{48}{(\tr A^2)}^4
               + \frac{121}{4}{(\tr A^2)}^2(\tr A^4)
               + \frac{19}{4}{(\tr A^4)}^2\\
           & & - \frac{158}{3}(\tr A^2)(\tr A^6),\nn\\
\tr F^{10} &=& - \frac{415}{96}{(\tr A^2)}^5
               + \frac{1985}{48}{(\tr A^2)}^3(\tr A^4)
               + \frac{45}{4}(\tr A^2){(\tr A^4)}^2\\
           & & - \frac{1805}{24}{(\tr A^2)}^2(\tr A^6)
               - \frac{5}{12}(\tr A^4)(\tr A^6).\nn
\end{eqnarray}%
\end{mathletters}%
The simplest relations involving the $\tr F^{2k}$ alone are these:
\begin{mathletters}
\begin{eqnarray}
\tr F^8    &=&   \frac{8683}{150000}{(\tr F^2)}^4
               - \frac{543}{500}{(\tr F^2)}^2(\tr F^4)
               + \frac{19}{4}{(\tr F^4)}^2\\
           & & + \frac{158}{375}(\tr F^2)(\tr F^6),\nn\\
\tr F^{10} &=&   \frac{8003}{300000}{(\tr F^2)}^5
               - \frac{2987}{6000}{(\tr F^2)}^3(\tr F^4)
               + \frac{11}{5}(\tr F^2){(\tr F^4)}^2\\
           & & + \frac{367}{3000}{(\tr F^2)}^2(\tr F^6)
               - \frac{1}{60}(\tr F^4)(\tr F^6).\nn
\end{eqnarray}%
\end{mathletters}%

\subsubsection{The Lie algebra $b_4$}

\paragraph{The defining representation:}

In the defining representation $(1,0,0,0)[9]$, the polynomials $\tr A^k$,
$k\in\{2,4,6,8\}$, can be used as generators of the algebra of invariant
polynomials. The others can be expressed in terms of the primitive ones, \eg:
\begin{eqnarray}
\tr A^{10} &=& - \frac{1}{384}{(\tr A^2)}^5 
               + \frac{5}{96}{(\tr A^2)}^3(\tr A^4)
               - \frac{5}{32}(\tr A^2){(\tr A^4)}^2\\
           & & - \frac{5}{24}{(\tr A^2)}^2(\tr A^6)
               + \frac{5}{12}(\tr A^4)(\tr A^6)
               + \frac{5}{8}(\tr A^2)(\tr A^8).\nn
\end{eqnarray}
The characteristic polynomial of the matrix $A$ is
\begin{eqnarray}
\chi_A(t) &=& t^9 - \frac{1}{2}(\tr A^2)\,t^7 
            +\Bigl(\frac{1}{8}{(\tr A^2)}^2 
                   - \frac{1}{4}(\tr A^4)\Bigr)\,t^5\\
        & & +\Bigl(- \frac{1}{48}{(\tr A^2)}^3 
                   + \frac{1}{8}(\tr A^2)(\tr A^4)
                   - \frac{1}{6}(\tr A^6)\Bigr)\,t^3\nn\\
        & & +\Bigl(\frac{1}{384}{(\tr A^2)}^4 
                   - \frac{1}{32}{(\tr A^2)}^2(\tr A^4)
                   + \frac{1}{32}{(\tr A^4)}^2
                   + \frac{1}{12}(\tr A^2)(\tr A^6)\nn\\
        & &        - \frac{1}{8}(\tr A^8)\Bigr)\,t.\nn 
\end{eqnarray}

\paragraph{The adjoint representation:}

The traces in the adjoint representation $(0,1,0,0)[36]$ can be expressed in
terms of the primitive polynomials via
\begin{mathletters}
\begin{eqnarray}
\tr F^2    &=&   7(\tr A^2),\\
\tr F^4    &=&   3{(\tr A^2)}^2 + (\tr A^4),\\
\tr F^6    &=&   15(\tr A^2)(\tr A^4) - 23(\tr A^6),\\
\tr F^8    &=&   35{(\tr A^4)}^2
               + 28(\tr A^2)(\tr A^6) - 119(\tr A^8),\\
\tr F^{10} &=&   \frac{503}{384}{(\tr A^2)}^5
               - \frac{2515}{96}{(\tr A^2)}^3(\tr A^4)
               + \frac{2515}{32}(\tr A^2){(\tr A^4)}^2\\
           & & + \frac{2515}{24}{(\tr A^2)}^2(\tr A^6)
               + \frac{5}{12}(\tr A^4)(\tr A^6)
               - \frac{2155}{8}(\tr A^2)(\tr A^8).\nn
\end{eqnarray}
\end{mathletters}
The simplest relation involving the $\tr F^{2k}$ alone is:
\begin{eqnarray}
\tr F^{10} &=& - \frac{657127}{2523470208}{(\tr F^2)}^5
               + \frac{2285}{262752}{(\tr F^2)}^3(\tr F^4)\\
           & & - \frac{4535}{87584}(\tr F^2){(\tr F^4)}^2
               - \frac{16385}{459816}{(\tr F^2)}^2(\tr F^6)\nn\\
           & & - \frac{5}{276}(\tr F^4)(\tr F^6)
               + \frac{2155}{6664}(\tr F^2)(\tr F^8).\nn
\end{eqnarray}

\subsection{The simple Lie algebras $c_\ell$}
\label{subapp_cn}

We define the matrices of the defining representations of the Lie algebras
$c_\ell$ by the matrices $A=a_jx_j$ given in section~\ref{subsect_cn}.  The
adjoint representations are defined by ${(F_j)}_{k\ell}=i\,c_{j\ell k}$ (where
$[x_i,x_j]=ic_{jk\ell}x_\ell$). The odd traces $\tr A^{2k-1}$ and $\tr
F^{2k-1}$, $k\in\N$, vanish.

\subsubsection{The Lie algebra $c_2$}

\paragraph{The defining representation:}

In the defining representation $(1,0)[4]$, the polynomials $\tr A^2$ and $\tr
A^4$ can be used as generators of the algebra of invariant polynomials.  The
others can be expressed in terms of the primitive ones:
\begin{mathletters}
\begin{eqnarray}
\tr A^6    &=& - \frac{1}{8}{(\tr A^2)}^3 + \frac{3}{4}(\tr A^2)(\tr A^4),\\
\tr A^8    &=& - \frac{1}{16}{(\tr A^2)}^4
               + \frac{1}{4}{(\tr A^2)}^2(\tr A^4)
               + \frac{1}{4}{(\tr A^4)}^2,\\
\tr A^{10} &=& - \frac{1}{64}{(\tr A^2)}^5 
               + \frac{5}{16}(\tr A^2){(\tr A^4)}^2,\\
\tr A^{12} &=& - \frac{3}{64}{(\tr A^2)}^4(\tr A^4)
               + \frac{3}{16}{(\tr A^2)}^2{(\tr A^4)}^2
               + \frac{1}{16}{(\tr A^4)}^3.
\end{eqnarray}%
\end{mathletters}%
The characteristic polynomial of the matrix $A$ is
\begin{equation}
\chi_A(t) = t^4 - \frac{1}{2}(\tr A^2)\,t^2 
  + \Bigl(\frac{1}{8}{(\tr A^2)}^2 - \frac{1}{4}(\tr A^4)\Bigr).
\end{equation}

\paragraph{The adjoint representation:}

The traces in the adjoint representation $(2,0)[10]$ can be expressed in terms
of the primitive polynomials via
\begin{mathletters}
\begin{eqnarray}
\tr F^2    &=& 6(\tr A^2),\\
\tr F^4    &=& 3{(\tr A^2)}^2 + 12(\tr A^4),\\
\tr F^6    &=& - \frac{9}{2}{(\tr A^2)}^3
               + 42(\tr A^2)(\tr A^4),\\
\tr F^8    &=& - \frac{47}{4}{(\tr A^2)}^4
               + 54{(\tr A^2)}^2(\tr A^4)
               + 68{(\tr A^4)}^2,\\
\tr F^{10} &=& - \frac{87}{8}{(\tr A^2)}^5
               - 15{(\tr A^2)}^3(\tr A^4)
               + 330(\tr A^2){(\tr A^4)}^2,\\
\tr F^{12} &=&   \frac{99}{16}{(\tr A^2)}^6
               - \frac{855}{4}{(\tr A^2)}^4(\tr A^4)
               + 789{(\tr A^2)}^2{(\tr A^4)}^2\\
           & & + 252{(\tr A^4)}^3.\nn
\end{eqnarray}%
\end{mathletters}%
The simplest relations involving the $\tr F^{2k}$ alone are these:
\begin{mathletters}
\begin{eqnarray}
\tr F^6    &=& - \frac{5}{72}{(\tr F^2)}^3 
               + \frac{7}{12}(\tr F^2)(\tr F^4),\\
\tr F^8    &=& - \frac{7}{432}{(\tr F^2)}^4
               + \frac{5}{108}{(\tr F^2)}^2(\tr F^4)
               + \frac{17}{36}{(\tr F^4)}^2,\\
\tr F^{10} &=&   \frac{1}{576}{(\tr F^2)}^5
               - \frac{5}{72}{(\tr F^2)}^3(\tr F^4)
               + \frac{55}{144}(\tr F^2){(\tr F^4)}^2,\\
\tr F^{12} &=&   \frac{35}{15552}{(\tr F^2)}^6
               - \frac{187}{5184}{(\tr F^2)}^4(\tr F^4)
               + \frac{25}{216}{(\tr F^2)}^2{(\tr F^4)}^2\\
           & & + \frac{7}{48}{(\tr F^4)}^3.\nn
\end{eqnarray}%
\end{mathletters}%

\subsubsection{The Lie algebra $c_3$}

\paragraph{The defining representation:}

In the defining representation $(1,0,0)[6]$, the polynomials $\tr A^2$, $\tr
A^4$ and $\tr A^6$ can be used as generators of the algebra of invariant
polynomials.  The others can be expressed in terms of the primitive ones:
\begin{mathletters}
\begin{eqnarray}
\tr A^8    &=&   \frac{1}{48}{(\tr A^2)}^4
               - \frac{1}{4}{(\tr A^2)}^2(\tr A^4)
               + \frac{1}{4}{(\tr A^4)}^2
               + \frac{2}{3}(\tr A^2)(\tr A^6),\\
\tr A^{10} &=&   \frac{1}{96}{(\tr A^2)}^5
               - \frac{5}{48}{(\tr A^2)}^3(\tr A^4)
               + \frac{5}{24}{(\tr A^2)}^2(\tr A^6)\\
           & & + \frac{5}{12}(\tr A^4)(\tr A^6).\nn
\end{eqnarray}%
\end{mathletters}%
The characteristic polynomial of the matrix $A$ is
\begin{eqnarray}
\chi_A(t) &=& t^6 -\frac{1}{2}(\tr A^2)\,t^4 
              + \Bigl(\frac{1}{8}{(\tr A^2)}^2 - \frac{1}{4}(\tr A^4)\Bigr)\,t^2\\
          &+& \Bigl(-\frac{1}{48}{(\tr A^2)}^3 
                + \frac{1}{8}(\tr A^2)(\tr A^4)
                - \frac{1}{6}(\tr A^6)\Bigr).\nn
\end{eqnarray}

\paragraph{The adjoint representation:}

The traces in the adjoint representation $(2,0,0)[21]$ can be expressed in
terms of the primitive polynomials via
\begin{mathletters}
\begin{eqnarray}
\tr F^2    &=&   8(\tr A^2),\\
\tr F^4    &=&   3{(\tr A^2)}^2 + 14(\tr A^4),\\
\tr F^6    &=&   15(\tr A^2)(\tr A^4) + 38(\tr A^6),\\
\tr F^8    &=&   \frac{67}{24}{(\tr A^2)}^4
               - \frac{67}{2}{(\tr A^2)}^2(\tr A^4)
               + \frac{137}{2}{(\tr A^4)}^2\\
           & & + \frac{352}{3}(\tr A^2)(\tr A^6),\nn\\
\tr F^{10} &=&   \frac{19}{3}{(\tr A^2)}^5
               - \frac{1565}{24}{(\tr A^2)}^3(\tr A^4)
               + \frac{45}{4}(\tr A^2){(\tr A^4)}^2\\
           & & + \frac{1655}{12}{(\tr A^2)}^2(\tr A^6)
               + \frac{2555}{6}(\tr A^4)(\tr A^6).\nn
\end{eqnarray}%
\end{mathletters}%
The simplest relations involving the $\tr F^{2k}$ alone are these:
\begin{mathletters}
\begin{eqnarray}
\tr F^8    &=&   \frac{2011}{357504}{(\tr F^2)}^4
               - \frac{1815}{14896}{(\tr F^2)}^2(\tr F^4)
               + \frac{137}{392}{(\tr F^4)}^2\\
           & & + \frac{22}{57}(\tr F^2)(\tr F^6),\nn\\
\tr F^{10} &=&   \frac{1081}{1430016}{(\tr F^2)}^5
               - \frac{2615}{357504}{(\tr F^2)}^3(\tr F^4)
               - \frac{745}{7448}(\tr F^2){(\tr F^4)}^2\\
           & & + \frac{35}{1824}{(\tr F^2)}^2(\tr F^6)
               + \frac{365}{456}(\tr F^4)(\tr F^6).\nn
\end{eqnarray}%
\end{mathletters}%

\subsubsection{The Lie algebra $c_4$}

\paragraph{The defining representation:}

In the defining representation $(1,0,0,0)[8]$, the polynomials $\tr A^k$,
$k\in\{2,4,6,8\}$ can be used as generators of the algebra of invariant
polynomials.  The others can be expressed in terms of the primitive ones, \eg
\begin{eqnarray}
\tr A^{10} &=& - \frac{1}{384}{(\tr A^2)}^5
               + \frac{5}{96}{(\tr A^2)}^3(\tr A^4)
               - \frac{5}{32}(\tr A^2){(\tr A^4)}^2\\
           & & - \frac{5}{24}{(\tr A^2)}^2(\tr A^6)
               + \frac{5}{12}(\tr A^4)(\tr A^6)
               + \frac{5}{8}(\tr A^2)(\tr A^8).\nn
\end{eqnarray}
The characteristic polynomial of the matrix $A$ is
\begin{eqnarray}
  \chi(t) &=& t^8 - \frac{1}{2}(\tr A^2)\,t^6 
              + \Bigl(\frac{1}{8}{(\tr A^2)}^2 
                - \frac{1}{4}(\tr A^4)\Bigr)\,t^4\\
          &+& \Bigl(-\frac{1}{48}{(\tr A^2)}
                + \frac{1}{8}(\tr A^2)(\tr A^4)
                - \frac{1}{6}(\tr A^6)\Bigr)\,t^2\nn\\
          &+& \Bigl(\frac{1}{384}{(\tr A^2)}^4 
                - \frac{1}{32}{(\tr A^2)}^2(\tr A^4)
                + \frac{1}{32}{(\tr A^4)}^2
              + \frac{1}{12}(\tr A^2)(\tr A^6)
                - \frac{1}{8}(\tr A^8)\Bigr).\nn
\end{eqnarray}

\paragraph{The adjoint representation:}

Let ${(F_j)}_{k\ell}=i\,c_{j\ell k}$ be the matrices of the adjoint
representation $(2,0,0,0)[36]$. The odd traces $\tr F^{2k-1}$, $k\in\N$,
vanish. The others can be expressed in terms of the primitive
polynomials via
\begin{mathletters}
\begin{eqnarray}
\tr F^2    &=& 10(\tr A^2),\\
\tr F^4    &=& 3{(\tr A^2)}^2 + 16(\tr A^4),\\
\tr F^6    &=& 15(\tr A^2)(\tr A^4) + 40(\tr A^6),\\
\tr F^8    &=& 35{(\tr A^4)}^2 + 28(\tr A^2)(\tr A^6) + 136(\tr A^8),\\
\tr F^{10} &=& - \frac{65}{48}{(\tr A^2)}^5
               + \frac{325}{12}{(\tr A^2)}^3(\tr A^4)
               - \frac{325}{4}(\tr A^2){(\tr A^4)}^2\\
           & & - \frac{325}{3}{(\tr A^2)}^2(\tr A^6)
               + \frac{1280}{3}(\tr A^4)(\tr A^6)
               + 370(\tr A^2)(\tr A^8).\nn
\end{eqnarray}
\end{mathletters}
The simplest relation involving the $\tr F^{2k}$ alone is:
\begin{eqnarray}
\tr F^{10} &=& - \frac{2039}{6528000}{(\tr F^2)}^5
               + \frac{2269}{163200}{(\tr F^2)}^3(\tr F^4)
               - \frac{143}{1088}(\tr F^2){(\tr F^4)}^2\\
           & & - \frac{1349}{20400}{(\tr F^2)}^2(\tr F^6)
               + \frac{2}{3}(\tr F^4)(\tr F^6)
               + \frac{37}{136}(\tr F^2)(\tr F^8).\nn
\end{eqnarray}

\subsection{The simple Lie algebras $d_\ell$}

We define the matrices of the defining representations of the Lie algebras
$d_\ell$ by the matrices $A=a_jx_j$ given in section~\ref{subsect_bn}.  In
order to generate the invariant polynomials of a simple Lie algebra of type
$d_\ell$, we need a square root of $\det(A)$ in addition to the trace
polynomials. The square of this invariant naturally appears
in~(\ref{char_general}) and is related to a Pfaffian form~\cite{Bo91,AzIz95}.

The adjoint representations are defined by ${(F_j)}_{k\ell}=i\,c_{j\ell k}$
(where $[x_i,x_j]=ic_{jk\ell}x_\ell$). The odd traces $\tr A^{2k-1}$ and $\tr
F^{2k-1}$, $k\in\N$, vanish.

\subsubsection{The Lie algebra $d_3$}

\paragraph{The defining representation:}

In the defining representation $(1,0,0)[6]$, the polynomials $\tr A^2$, $\tr
A^4$ and $\sqrt{\det A}$, which is of degree three, can be used as generators
of the algebra of invariant polynomials. The others can be expressed in terms
of the primitive ones, \eg:
\begin{mathletters}
\begin{eqnarray}
\tr A^6    &=& - \frac{1}{8}{(\tr A^2)}^3 - 6(\det A) 
               + \frac{3}{4}(\tr A^2)(\tr A^4),\\
\tr A^8    &=& - \frac{1}{16}{(\tr A^2)}^4
               - 4(\tr A^2)(\det A)
               + \frac{1}{4}{(\tr A^2)}^2(\tr A^4)
               + \frac{1}{4}{(\tr A^4)}^2,\\
\tr A^{10} &=& - \frac{1}{64}{(\tr A^2)}^5
               - \frac{5}{4}{(\tr A^2)}^2(\det A)
               - \frac{5}{2}(\det A)(\tr A^4)\\
           & & + \frac{5}{16}(\tr A^2){(\tr A^4)}^2.\nn
\end{eqnarray}
\end{mathletters}
The characteristic polynomial of the matrix $A$ is
\begin{equation}
\chi(t) = t^6 - \frac{1}{2}(\tr A^2)\,t^4 
            + \Bigl(\frac{1}{8}{(\tr A^2)}^2 - \frac{1}{4}(\tr A^4)\Bigr)\,t^2\\ 
            + (\det A).
\end{equation}

\paragraph{The adjoint representation:}

The traces in the adjoint representation $(0,1,1)[15]$ can be expressed in
terms of the primitive polynomials via
\begin{mathletters}
\begin{eqnarray}
\tr F^2    &=&   4(\tr A^2),\\
\tr F^4    &=&   3{(\tr A^2)}^2 - 2(\tr A^4),\\
\tr F^6    &=&   \frac{13}{4}{(\tr A^2)}^3
               + 156(\det A)
               - \frac{9}{2}(\tr A^2)(\tr A^4),\\
\tr F^8    &=&   \frac{33}{8}{(\tr A^2)}^4
               + 320(\tr A^2)(\det A)
               - \frac{19}{2}{(\tr A^2)}^2(\tr A^4)
               + \frac{9}{2}{(\tr A^4)}^2,\\
\tr F^{10} &=&   \frac{163}{32}{(\tr A^2)}^5
               + \frac{905}{2}{(\tr A^2)}^2(\det A)
               - 15{(\tr A^2)}^3(\tr A^4)\\
           & & + 5(\det A)(\tr A^4)
               + \frac{85}{8}(\tr A^2){(\tr A^4)}^2.\nn
\end{eqnarray}
\end{mathletters}
The simplest relations involving the $\tr F^{2k}$ alone are
\begin{mathletters}
\begin{eqnarray}
\tr F^8    &=&   \frac{35}{1248}{(\tr F^2)}^4
               - \frac{43}{104}{(\tr F^2)}^2(\tr F^4)
               + \frac{9}{8}{(\tr F^4)}^2\\
           & & + \frac{20}{39}(\tr F^2)(\tr F^6),\nn\\
\tr F^{10} &=&   \frac{41}{2496}{(\tr F^2)}^5
               - \frac{295}{1248}{(\tr F^2)}^3(\tr F^4)
               + \frac{35}{52}(\tr F^2){(\tr F^4)}^2\\
           & & + \frac{115}{624}{(\tr F^2)}^2(\tr F^6)
               - \frac{5}{312}(\tr F^4)(\tr F^6).\nn
\end{eqnarray}
\end{mathletters}

\subsubsection{The Lie algebra $d_4$}

\paragraph{The defining representation:}

In the defining representation $(1,0,0,0)[8]$, the polynomials $\tr
A^k$, $k\in\{2,4,6\}$, and $\sqrt{\det A}$, which, like $\tr A^4$, is
of degree four, can be used as generators of the algebra of invariant
polynomials.  The others can be expressed in terms of the primitive
ones, \eg:
\begin{mathletters}
\begin{eqnarray}
\tr A^8    &=&   \frac{1}{48}{(\tr A^2)}^4 
               - 8(\det A)
               - \frac{1}{4}{(\tr A^2)}^2(\tr A^4)
               + \frac{1}{4}{(\tr A^4)}^2\\
           & & + \frac{2}{3}(\tr A^2)(\tr A^6),\nn\\
\tr A^{10} &=&   \frac{1}{96}{(\tr A^2)}^5 
               - 5(\tr A^2)(\det A)
               - \frac{5}{48}{(\tr A^2)}^3(\tr A^4)\\
           & & + \frac{5}{24}{(\tr A^2)}^2(\tr A^6)
               + \frac{5}{12}(\tr A^4)(\tr A^6).\nn
\end{eqnarray}%
\end{mathletters}%
The characteristic polynomial of the matrix $A$ is
\begin{eqnarray}
\chi(t) &=& t^8 - \frac{1}{2}(\tr A^2)\,t^6 
            + \Bigl(\frac{1}{8}{(\tr A^2)}^2 - \frac{1}{4}(\tr A^4)\Bigr)\,t^4 \\
        &+& \Bigl( -\frac{1}{48}{(\tr A^2)}^3 
            + \frac{1}{8}(\tr A^2)(\tr A^4)
            - \frac{1}{6}(\tr A^6)\Bigr)\,t^2
         +  (\det A).\nn
\end{eqnarray}

\paragraph{The adjoint representation:}

The traces in the adjoint representation $(0,1,0,0)[28]$ can be expressed in
terms of the primitive polynomials via
\begin{mathletters}
\begin{eqnarray}
\tr F^2    &=&   6(\tr A^2),\\
\tr F^4    &=&   3{(\tr A^2)}^2,\\
\tr F^6    &=&   15(\tr A^2)(\tr A^4) -24(\tr A^6),\\
\tr F^8    &=& - \frac{5}{2}{(\tr A^2)}^4
               + 960(\det A)
               + 30{(\tr A^2)}^2(\tr A^4)\\
           & & + 5{(\tr A^4)}^2
               - 52(\tr A^2)(\tr A^6),\nn\\
\tr F^{10} &=& - \frac{69}{16}{(\tr A^2)}^5
               + 2160(\tr A^2)(\det A)
               + \frac{165}{4}{(\tr A^2)}^3(\tr A^4)\\
           & & + \frac{45}{4}(\tr A^2){(\tr A^4)}^2
               - 75{(\tr A^2)}^2(\tr A^6).\nn
\end{eqnarray}%
\end{mathletters}%
The simplest relations involving the $\tr F^{2k}$ alone are
\begin{mathletters}
\begin{eqnarray}
\tr F^{4}  &=&   \frac{1}{12}{(\tr F^2)}^2,\\
\tr F^{10} &=&   \frac{7}{41472}{(\tr F^2)}^5
               - \frac{7}{144}{(\tr F^2)}^2(\tr F^6)
               + \frac{3}{8}(\tr F^2)(\tr F^8),\\
\tr F^{14} &=& - \frac{2761}{179159040}{(\tr F^2)}^7
               + \frac{24409}{5598720}{(\tr F^2)}^4(\tr F^6)\\
           & & - \frac{1001}{19440}(\tr F^2){(\tr F^6)}^2
               - \frac{7931}{311040}{(\tr F^2)}^3(\tr F^8)\nn\\
           & & + \frac{77}{360}(\tr F^6)(\tr F^8)
               + \frac{497}{1080}(\tr F^2)(\tr F^{12}).\nn
\end{eqnarray}%
\end{mathletters}%

\subsection{The simple Lie algebra $e_6$}
\label{subapp_e6}

\paragraph{The defining representation:}

We denote the matrices of the defining representation $(1,0,0,0,0,0)[27]$ of
$e_6$ by $A$. We have $\tr A=0$ and $\tr A^3=0$, and the polynomials $\tr
A^k$, $k\in\{2,5,6,8,9,12\}$ are suitable generators of the algebra of
invariant polynomials. The relations are
\begin{mathletters}
\begin{eqnarray}
\tr A^4    &=&   \frac{1}{12}{(\tr A^2)}^2,\\
\tr A^7    &=&   \frac{7}{24}(\tr A^2)(\tr A^5),\\
\tr A^{10} &=&   \frac{7}{41472}{(\tr A^2)}^5
               + \frac{7}{40}{(\tr A^5)}^2
               - \frac{7}{144}{(\tr A^2)}^2(\tr A^6)\\
           & & + \frac{3}{8}(\tr A^2)(\tr A^8),\nn\\
\tr A^{11} &=& - \frac{55}{3456}{(\tr A^2)}^3(\tr A^5) 
               + \frac{11}{36}(\tr A^5)(\tr A^6)
               + \frac{605}{1512}(\tr A^2)(\tr A^9),\\
\tr A^{13} &=& - \frac{143}{27648}{(\tr A^2)}^4(\tr A^5)
               + \frac{143}{2700}(\tr A^2)(\tr A^5)(\tr A^6)\\
           & & + \frac{143}{400}(\tr A^5)(\tr A^8)
               + \frac{1859}{18144}{(\tr A^2)}^2(\tr A^9).\nn
\end{eqnarray}%
\end{mathletters}%

\paragraph{The adjoint representation:}

Let ${(F_j)}_{k\ell}=i\,c_{j\ell k}$ be the matrices of the adjoint
representation $(0,0,0,0,0,1)[78]$. The odd traces $\tr F^{2k-1}$,
$k\in\N$, vanish. The others can be expressed in terms of the primitive
polynomials via
\begin{mathletters}
\label{tracee6_fa}
\begin{eqnarray}
\tr F^2    &=&   4(\tr A^2),\\
\tr F^4    &=&   \frac{1}{2}{(\tr A^2)}^2,\\
\tr F^6    &=&   \frac{5}{36}{(\tr A^2)}^3 - 6(\tr A^6),\\
\tr F^8    &=&   \frac{35}{432}{(\tr A^2)}^4
               - \frac{28}{3}(\tr A^2)(\tr A^6)
               + 18(\tr A^8),\\
\tr F^{10} &=&   \frac{91}{2304}{(\tr A^2)}^5
               - \frac{21}{20}{(\tr A^5)}^2
               - \frac{133}{24}{(\tr A^2)}^2(\tr A^6)\\
           & & + \frac{51}{4}(\tr A^2)(\tr A^8).\nn
\end{eqnarray}%
\end{mathletters}%

\subsection{The simple Lie algebra $f_4$}
\label{subapp_f4}

\paragraph{The defining representation:}

We denote the matrices of the defining representation $(0,0,0,1)[26]$
of $f_4$ by $A$. The polynomials $\tr A^k$, $k\in\{2,6,8,12\}$ are
suitable generators of the algebra of invariant polynomials. The odd
traces $\tr A^{2k-1}$, $k\in\N$, vanish. The relations are
\begin{mathletters}
\begin{eqnarray}
\tr A^4    &=&   \frac{1}{12}{(\tr A^2)}^2,\\
\tr A^{10} &=&   \frac{7}{41472}{(\tr A^2)}^5 
               - \frac{7}{144}{(\tr A^2)}^2(\tr A^6)
               + \frac{3}{8}(\tr A^2)(\tr A^8),\\
\tr A^{14} &=& - \frac{2761}{179159040}{(\tr A^2)}^7 
               + \frac{24409}{5598720}{(\tr A^2)}^4(\tr A^6)\\
           & & - \frac{1001}{19440}(\tr A^2){(\tr A^6)}^2
               - \frac{7931}{311040}{(\tr A^2)}^3(\tr A^8)\nn\\
           & & + \frac{77}{360}(\tr A^6)(\tr A^8)
               + \frac{497}{1080}(\tr A^2)(\tr A^{12}).\nn
\end{eqnarray}%
\end{mathletters}%

\paragraph{The adjoint representation:}

Let ${(F_j)}_{k\ell}=i\,c_{j\ell k}$ be the matrices of the adjoint
representation $(1,0,0,0)[52]$ in a suitable normalization relative to
those of the defining representation. The odd traces $\tr F^{2k-1}$,
$k\in\N$, vanish. The others can be expressed in terms of the primitive
polynomials via
\begin{mathletters}
\begin{eqnarray}
\tr F^2    &=&   3(\tr A^2),\\
\tr F^4    &=&   \frac{5}{12}{(\tr A^2)}^2,\\
\tr F^6    &=&   \frac{5}{36}{(\tr A^2)}^3 - 7(\tr A^6),\\
\tr F^8    &=&   \frac{35}{432}{(\tr A^2)}^4 
               - \frac{28}{3}(\tr A^2)(\tr A^6) + 17(\tr A^8),\\
\tr F^{10} &=&   \frac{1631}{41472}{(\tr A^2)}^5 
               - \frac{791}{144}{(\tr A^2)}^2(\tr A^6) 
               + \frac{99}{8}(\tr A^2)(\tr A^8),\\
\tr F^{12} &=&   \frac{1309}{62208}{(\tr A^2)}^6
               - \frac{2387}{648}{(\tr A^2)}^3(\tr A^6)
               + \frac{154}{9}{(\tr A^6)}^2\\
           & & + \frac{209}{18}{(\tr A^2)}^2(\tr A^8)
               - 63 (\tr A^{12}).\nn
\end{eqnarray}%
\end{mathletters}%
The simplest relations of the $\tr F^{2k}$ are
\begin{mathletters}
\label{tracef4_ff}
\begin{eqnarray}
\tr F^4    &=&   \frac{5}{108}{(\tr F^2)}^2,\\
\label{tracef4_aab}
\tr F^{10} &=&   \frac{161}{6345216}{(\tr F^2)}^5
               - \frac{455}{22032}{(\tr F^2)}^2(\tr F^6)
               + \frac{33}{136}(\tr F^2)(\tr F^8).
\end{eqnarray}%
\end{mathletters}%

\subsection{The simple Lie algebra $g_2$}
\label{subapp_g2}

\paragraph{The defining representation:}

We define the defining representation $(0,1)[7]$ by the same matrices as in
section~\ref{subsect_g2}, $A=a_jx_j$.  The polynomials $\tr A^2$ and $\tr A^6$
can be taken as generators of the algebra of invariant polynomials. The odd
traces $\tr A^{2k-1}$, $k\in\N$, vanish. The $\tr A^k$ satisfy these relations:
\begin{mathletters}
\begin{eqnarray}
\tr A^4    &=&   \frac{1}{4}{(\tr A^2)}^2,\\
\tr A^8    &=& - \frac{5}{192}{(\tr A^2)}^4 + \frac{2}{3}(\tr A^2)(\tr A^6),\\
\tr A^{10} &=& - \frac{1}{64}{(\tr A^2)}^5 + \frac{5}{16}{(\tr A^2)}^2(\tr A^6),\\
\tr A^{12} &=& - \frac{19}{3072}{(\tr A^2)}^6 
               + \frac{5}{48}{(\tr A^2)}^3(\tr A^6)
               + \frac{1}{6}{(\tr A^6)}^2.
\end{eqnarray}%
\end{mathletters}%
The characteristic polynomial of the matrix $A$ is
\begin{eqnarray}
\chi_A(t) &=& t^7 - \frac{1}{2}(\tr A^2)\,t^5 
  + \frac{1}{16}{(\tr A^2)}^2\,t^3
  + \Bigl(\frac{1}{96}{(\tr A^2)}^3 - \frac{1}{6}(\tr A^6)\Bigr)\,t.
\end{eqnarray}

\paragraph{The adjoint representation:}

Let ${(F_j)}_{k\ell} = i\,c_{j\ell k}$ be the matrices of the adjoint
representation $(1,0)[14]$ (where $[x_j,x_k]=i\,c_{jk\ell}x_\ell$). The odd
traces $\tr F^{2k-1}$, $k\in\N$, vanish. The others can be expressed in
terms of the
primitive polynomials via
\begin{mathletters}
\begin{eqnarray}
\label{traceg2_faa}
\tr F^2    &=&   4(\tr A^2),\\
\tr F^4    &=&   \frac{5}{2}{(\tr A^2)}^2,\\
\tr F^6    &=&   \frac{15}{4}{(\tr A^2)}^3 -26(\tr A^6),\\
\label{traceg2_fad}
\tr F^8    &=&   \frac{515}{96}{(\tr A^2)}^4 -\frac{160}{3}(\tr A^2)(\tr A^6),\\
\tr F^{10} &=&   \frac{431}{64}{(\tr A^2)}^5 -\frac{605}{8}{(\tr A^2)}^2(\tr A^6),\\
\tr F^{12} &=&   \frac{12865}{1536}{(\tr A^2)}^6 
               - \frac{1315}{12}{(\tr A^2)}^3(\tr A^6)
               + \frac{365}{3}{(\tr A^6)}^2.
\end{eqnarray}%
\end{mathletters}%
The $\tr F^{2k}$ themselves satisfy these relations:
\begin{mathletters}
\begin{eqnarray}
\label{traceg2_ffa}
\tr F^4    &=&   \frac{5}{32}{(\tr F^2)}^2,\\
\label{traceg2_ffb}
\tr F^8    &=& - \frac{2905}{319488}{(\tr F^2)}^4 + \frac{20}{39}(\tr F^2)(\tr F^6),\\
\tr F^{10} &=& - \frac{217}{53248}{(\tr F^2)}^5 
               + \frac{605}{3328}{(\tr F^2)}^2(\tr F^6).
\end{eqnarray}%
\end{mathletters}%
The characteristic polynomial of the matrix $F$ is
\begin{mathletters}
\begin{eqnarray}
\chi_F(t) &=& t^{14}
            - 2(\tr A^2)\,t^{12}
            + \frac{11}{8}{(\tr A^2)}^2\,t^{10}
            + \Bigl(-\frac{17}{24}{(\tr A^2)}^3 + \frac{13}{3}(\tr A^6)\Bigr)\,t^8\\
        & & + \Bigl(\frac{49}{256}{(\tr A^2)}^4 -2(\tr A^2)(\tr A^6)\Bigr)\,t^6\nn\\
        & & + \Bigl(-\frac{1}{64}{(\tr A^2)}^5 
              + \frac{3}{16}{(\tr A^2)}^2(\tr A^6)\Bigr)\,t^4\nn\\ 
        & & + \Bigl(-\frac{11}{3072}{(\tr A^2)}^6 
              + \frac{5}{48}{(\tr A^2)}^3(\tr A^6)
              - \frac{3}{4}{(\tr A^6)}^2\Bigr)\,t^2\nn\\
        &=& t^{14}
            -\frac{1}{2}(\tr F^2)\,t^{12}
            +\frac{11}{128}{(\tr F^2)}^2\,t^{10}\\
        & & + \Bigl(-\frac{1}{768}{(\tr F^2)}^3 -\frac{1}{6}(\tr F^6)\Bigr)\,t^8\nn\\
        & & + \Bigl(-\frac{323}{851968}{(\tr F^2)}^4 
            + \frac{1}{52}(\tr F^2)(\tr F^6)\Bigr)\,t^6\nn\\
        & & + \Bigl(\frac{19}{1703936}{(\tr F^2)}^5 
            - \frac{3}{6656}{(\tr F^2)}^2(\tr F^6)\Bigr)\,t^4\nn\\
        & & + \Bigl(-\frac{2159}{2126512128}{(\tr F^2)}^6 
            + \frac{35}{519168}{(\tr F^2)}^3(\tr F^6)
            - \frac{3}{2704}{(\tr F^6)}^2\Bigr)\,t^2.\nn
\end{eqnarray}%
\end{mathletters}%


\end{document}